\documentclass{ws-ijbc}
\usepackage{ws-rotating}     

\usepackage{graphicx,color}
\usepackage[colorlinks,linkcolor=blue,citecolor=blue,urlcolor=blue]{hyperref}

\begin{document}

\catchline{}{}{}{}{} 

\markboth{M. K. Krotter {\it et al.}}{CUTTING AND SHUFFLING A LINE SEGMENT}

\title{CUTTING AND SHUFFLING A LINE SEGMENT: MIXING BY INTERVAL EXCHANGE TRANSFORMATIONS}

\author{MARISSA K. KROTTER} 
\address{Department of Mechanical Engineering, Northwestern University,\\
Evanston, Illinois 60208, USA\\
MarissaKrotter2012@u.northwestern.edu} 

\author{IVAN C. CHRISTOV\footnote{Present address: Department of Mechanical and Aerospace Engineering, Princeton University, Princeton, NJ 08544, USA.}}
\address{Department of Engineering Sciences and Applied Mathematics, Northwestern University,\\
Evanston, Illinois 60208, USA\\
christov@u.northwestern.edu}

\author{JULIO M. OTTINO}
\address{Department of Chemical and Biological Engineering, and \\
The Northwestern Institute on Complex Systems (NICO), Northwestern University,\\
Evanston, Illinois 60208, USA\\
jm-ottino@northwestern.edu}

\author{RICHARD M. LUEPTOW\footnote{Author to whom correspondence should be addressed.}}
\address{Department of Mechanical Engineering, and \\
The Northwestern Institute on Complex Systems (NICO), Northwestern University,\\
Evanston, Illinois 60208, USA\\
r-lueptow@northwestern.edu}

\maketitle

\begin{history}
\received{(to be inserted by publisher)}
\end{history}

\begin{abstract}
We present a computational study of finite-time mixing of a line segment by cutting and shuffling. A family of one-dimensional interval exchange transformations is constructed as a model system in which to study these types of mixing processes. Illustrative examples of the mixing behaviors, including pathological cases that violate the assumptions of the known governing theorems and lead to poor mixing, are shown. Since the mathematical theory applies as the number of iterations of the map goes to infinity, we introduce practical measures of mixing (the percent unmixed and the number of intermaterial interfaces) that can be computed over given (finite) numbers of iterations. We find that good mixing can be achieved after a finite number of iterations of a one-dimensional cutting and shuffling map, even though such a map cannot be considered chaotic in the usual sense and/or it may not fulfill the conditions of the ergodic theorems for interval exchange transformations. Specifically, good shuffling can occur with only six or seven intervals of roughly the same length, as long as the rearrangement order is an irreducible permutation. This study has implications for a number of mixing processes in which discontinuities arise either by construction  or due to the underlying physics.
\end{abstract}

\keywords{Cutting and shuffling; piecewise isometries; granular mixing; lamellar structures.}


\section{Introduction}

The precise ``symptoms'' expressed by a chaotic dynamical system have been the subject of much discussion over the years \cite{cb96,cb98,mds98,cb99}. A positive leading Lyapunov exponent, computed analytically or from a time series, is one common requirement \cite[Chap.~5]{bg96}. Recently, piecewise-continuous and discontinuous dynamical systems have provided examples and counterexamples to the ``typical'' notions associated with chaos, often leading to the development of new types of bifurcation phenomena \cite{bbcknop08}, and even the possibility of  \emph{nondeterministic chaos} \cite{j11}. An overview of the role of discontinuities in mixing, illustrated by recent examples from the literature, is provided by \citet{sturman12}.

\emph{Piecewise isometries} (PWIs) \cite{g00,g02} are one of the most recent examples of a discontinuous dynamical system exhibiting nontrivial dynamics. Under some specific conditions, it can be shown that the topological entropy \cite{b01} and all of the Lyapunov exponents \cite{fd08} of a PWI are equal to zero (wherever and whenever they exist). A smooth distance-preserving map does not lead to complex dynamics in any sense of the word, but ``gluing'' together different isometries in a piecewise manner across a line of discontinuity leads to highly nontrivial behavior of orbits. The resulting complex dynamics include, but are not limited to, attractors and quasi-periodicity \cite{g98,shm01,k02,mn04,ag05,lv10}. In fact, \citet{k09,k09b} has shown that certain PWIs satisfy a modification of Devaney's criterion for chaos \cite[\S1.8]{d03}. A dynamical system is Devaney chaotic if it is (i) sensitive to initial conditions, (ii) topologically transitive and (iii) has a set of periodic orbits that is dense in the domain; planar PWIs satisfy criteria (i) and (ii).

PWIs also arise in the study of digital filters \cite{cl88,o92}. If one allows for the more general class of piecewise \emph{contraction} (non-area-preserving) maps, there are further applications to electrical engineering \cite{d06}. More recently, PWI maps have been proposed as the skeleton of the kinematics of the flow of granular materials in tumblers \cite{col10,col10b,jlosw10,jcol12}. The PWI dynamics in granular mixing are termed \emph{cutting and shuffling} to emphasize their fundamental difference from the well-known \emph{stretching and folding} mechanism of chaotic fluid mixing \cite{o89}. Some illustrated examples of these differences are provided by \citet{clo11}. 

Laboratory experiments have shown \cite{jlosw10,jcol12} that the mixing properties of granular flows can be predicted based on the underlying PWI framework. However, it is also worthwhile to consider simpler PWI-type mixing problems. The one-dimensional (1D) version of a PWI is an \emph{interval exchange transformation} (IET) \cite{g98}. In a 1D context, cutting and shuffling takes a particularly intuitive form, well known to card players \cite{g61,ad86,tt00}. Recent work by \citet{lw08} showing that 2D chaotic fluid mixing can exhibit \emph{cutoffs} (i.e., abrupt transitions from partially-mixed to well-mixed states) usually associated with card shuffling \cite{ad86,tt00} provides further evidence for the relevance of 1D cutting and shuffling. IETs, which replace the deck of cards with a set of  subsegments of variable lengths, are one of the simplest dynamical systems that can be successfully studied analytically in some detail \cite[\S14.5]{kh95}. More striking, however, is that IETs describe the dynamics along lines of discontinuity in some planar PWIs \cite{k09b}. 

\begin{figure}
\begin{center}
\includegraphics{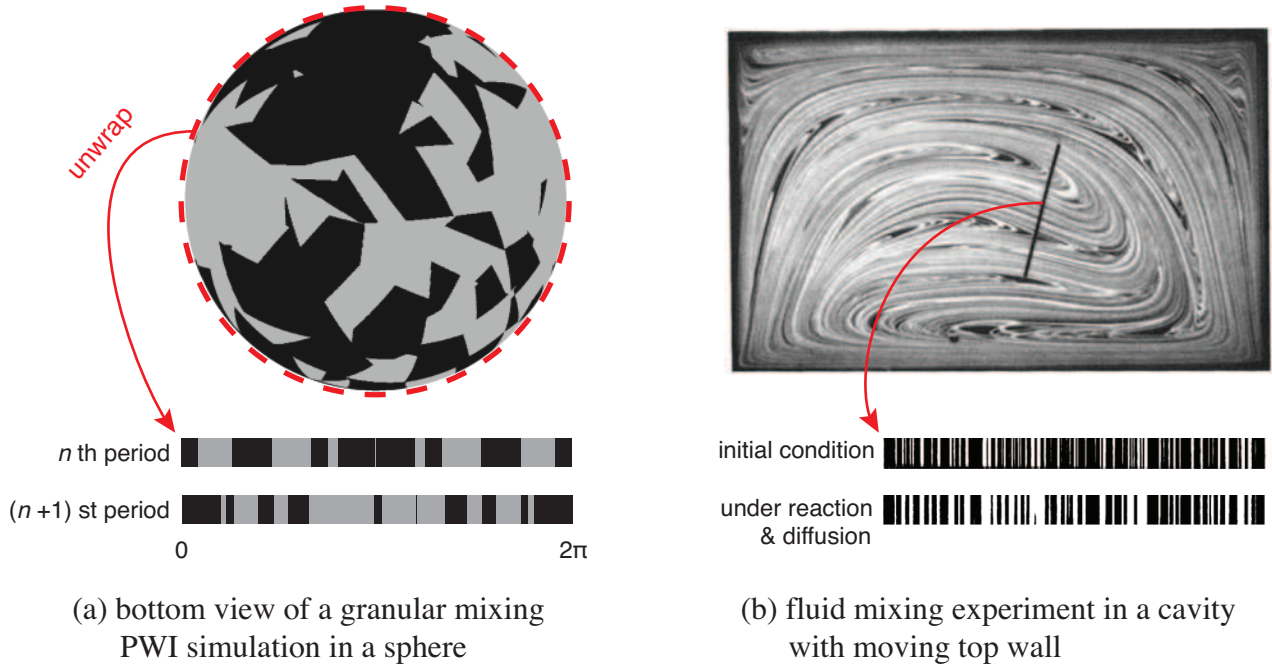}
\end{center}
\caption{(a) View from below of a PWI granular mixing simulation in a ``blinking spherical tumbler flow'' \cite{jlosw10,jcol12}. Initially, the black and gray materials occupy the left and right quarter-spheres, respectively. Inset shows the cutting and shuffling dynamics occurring at the discontinuity interface (dotted curve). Shuffling occurs upon rotation about an axis, while cutting results from changing the axis of rotation. (b) Photograph showing typical structures produced by chaotic mixing in a cavity flow experiment using two fluids of similar viscosity and negligible interfacial tension. A lamellar structure with striations of distributed thickness is generated by the chaotic flow. Inset shows a cross-section, which is the starting point for a 1D  model of the evolution of the lamellar structure. The strips in the insets in panels (a) and (b) have a finite vertical extent for clarity of presentation. Images in panel (b) reproduced, with permission, from F.J.\ Muzzio \& J.M.\ Ottino, {\it Phys.\ Rev.\ Lett.}, {\bf 63}, 1989, 47--50. \copyright\ 1989 American Physical Society.}
\label{mk:fig:lamella}
\end{figure}

To illustrate mixing by cutting and shuffling in the context of granular flow, consider a spherical device partially-filled with two initially separate colors of granular material in which the mixing dynamics due to short rotations about two axes can be observed from below \cite{mlo07,smow08,jlosw10}. Figure~\ref{mk:fig:lamella}(a) shows a bottom view of a PWI simulation of such a setup after several iterations of these rotations. As the sphere rotates about an axis, material moves from one side of the surface in Fig.~\ref{mk:fig:lamella}(a) to the other, and re-enters the side it came from upon reaching the dashed edge. For the PWI framework to be valid, we must first assume that the re-entry occurs instantaneously \cite{col10,jlosw10}, i.e., there is a discontinuous mapping from one side of the dotted circle in Fig.~\ref{mk:fig:lamella}(a) to the other across the axis of rotation (not shown and arbitrary). IET-like dynamics are evident if we ``unwrap'' the boundary of the surface of discontinuity shown approximately by the dotted curve encircling the mixing pattern in Fig.~\ref{mk:fig:lamella}(a). The dynamics occurring on this interface are the source of complexity in the PWI model of granular mixing in a spherical tumbler \cite[\S8.9]{c11}. Thus, an understanding of the behavior of such IETs can shed light on the more complicated PWI dynamics in a plane or on a sphere. Similar discontinuous changes in the flow also occur in a number of chaotic fluid mixing problems such as the periodic reorientation of streamlines in a circular Hele-Shaw cell, which has been used as a model for injection wells in porous geological formations \cite{mlokthrlm10}. A detailed discussion of how such flows fit into the PWI framework is given by \citet{sturman12}.

The concept of intermaterial area is central in the theory of chaotic mixing \cite{o89}. A cross-section of the mixing process reveals a pattern of pieces of material of different types and varying thickness. Stretching and folding motions create fine lamellar structures, such as those seen in Fig.~\ref{mk:fig:lamella}(b), across which reaction and diffusion processes act. The distribution of the thickness of lamella along a 1D slice of the domain (e.g., across the black line in Fig.~\ref{mk:fig:lamella}(b)) has been studied, showing universal properties of the evolution of the distribution toward a steady state \cite{mo89}. One-dimensional models of the lamellar dynamics are useful, because precise numerical tracking of the intermaterial area in a chaotic flow, even in 2D, is a difficult (if not impossible) task \cite{fo87}. What is more, the effective 1D dynamics across a cross-section (as in Fig.~\ref{mk:fig:lamella}(b)) can be thought of as an IET with molecular diffusion blurring the boundaries between subsegments before they are permuted. Different models of ``cutting'' can lead to significant enhancement over a purely diffusive mixing process \cite{omtfjk92,ank02}. In an industrial setting, various designs of mixers have been proposed that create such lamellar structures in a controlled and reproducible fashion \cite{hm97,ndpm11}. In such a device, the speed of a two-step chemical reaction depends sensitively on the arrangement and thickness of the lamella \cite{ccr99,ccr00}. More importantly, in multiple reactions, the final product distribution depends on the dynamics of mixing \cite{co84}. Therefore, a better understanding of the practical details of IETs can lead to improvements in the design of devices that must mix materials or allow them to react completely in a precise manner.

While the mathematical theory of IETs is mature (see, e.g., the recent review of \citet{v06}), few computational examples and parameter space studies can be found in the literature. Moreover,   theorems focus on \emph{infinite-time} concepts such as ergodicity, weak mixing and Lyapunov exponents. It is apparent, however, that infinite-time concepts, while very useful, do not fully address the questions that arise in practice. Mixing experiments and chemical reactions take place over a \emph{finite} period of time. For mixing of granular materials, the diameter $d$ of a typical particle defines a ``cut off'' length scale beyond which  further mixing is impossible because smaller-scale structures cannot be produced. Similarly, in fluid mixing the diffusion length scale $\sim\sqrt{Dt}$, where $D$ is the molecular diffusivity and $t$ is time, sets a lower-bound beyond which it is not worthwhile to continue creating finer lamellar structures \cite{tc99}. Therefore, questions such as ``How many iterations does it take to reduce longest continuous segment in an IET to a given fraction of the interval?'' cannot be answered by current theory on IETs.

The only previous detailed numerical study of mixing of a line segment is that of \citet{ank02}, who only consider protocols based on exchanges of equal-length subsegments, which do not necessarily mix. \citet{ank02} noted that ``mixing properties of interval exchange maps are very subtle and relatively poorly understood and depend on parameters in a sensitive way.'' Thus, the goal of the present work is to extend the latter study to mixing by cutting and shuffling within the framework of the ergodic theory developed for interval exchange transformations (IETs). Specifically, we aim to develop illustrative examples of the behavior of 1D cutting and shuffling maps in order to clarify how the pathological cases that lead to poor mixing (e.g., those that violate the known IET theorems). We shall do so in the context of finite-time transport, characterizing the mixing properties of those maps over a given (fixed) number of iterations. The proofs of the ergodic and mixing theorems for IETs are difficult, and similar questions for PWIs remain open. As a result, computational studies have recently proven to be an important part of the mathematical discovery process in this field \cite{bp07}.

This paper is organized as follows. Section~\ref{mk:sec:IET_construct} introduces our IET construction and defines the parameter space to be explored. Section~\ref{mk:sec:theory} summarizes the relevant previous mathematical results regarding IETs, their complexity and their mathematical mixing properties. In Sec.~\ref{mk:sec:num_res}, we present practical numerical results on mixing by 1D cutting and shuffling, and connect the behaviors to the mathematical theory from Sec.~\ref{mk:sec:theory}. Section~\ref{mk:sec:conclusion} summarizes the most important results of our study and presents our conclusions.

\section{IET construction and simulation methodology}
\label{mk:sec:IET_construct}

\begin{figure}[h]
\begin{center}
\includegraphics{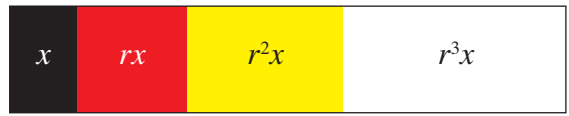}
\end{center}
\caption{(Color online.) Sketch of how an interval $\mathcal{I}$ is subdivided in our IET construction with $N = 4$ and $r = 1.5$.}
\label{mk:fig:ic}
\end{figure}

To study 1D cutting and shuffling, i.e., the ``mixing'' of a line segment $\mathcal{I}$, we construct a specific type of IET, which we denote as $T_{S,\Pi}$: $\mathcal{I}\to\mathcal{I}$, with a well-defined parameter space that we can systematically explore. To this end, without loss of generality, we take the line segment to be an interval of the real line, specifically $\mathcal{I}=[0,1]$. Then, cutting and shuffling proceeds by dividing $\mathcal{I}$ into a collection of $N$ disjoint subsegments given by the set $S = \{\mathcal{I}_1,\hdots,\mathcal{I}_N\}$. Specifically, the interval is divided into $N$ subsegments of length $|\mathcal{I}_i|$ and the ratio $r_i = |\mathcal{I}_{i}|/|\mathcal{I}_{i-1}|$ between the lengths of consecutive subsegments is taken to be constant (i.e., $r_i=r\ge1$ $\forall i$), as shown in Fig.~\ref{mk:fig:ic}. Now, we specify $r$ and require the length of the interval $\mathcal{I}$ to be 1, which gives
\begin{equation}
\sum_{i=1}^{N} r^{i-1} x = 1 \quad\Rightarrow\quad x = \frac{r-1}{r^N-1},
\label{mk:eq:x}
\end{equation}
where $x=|\mathcal{I}_1|$ is the primary interval length.

\begin{figure}[h]
\begin{center}
\includegraphics{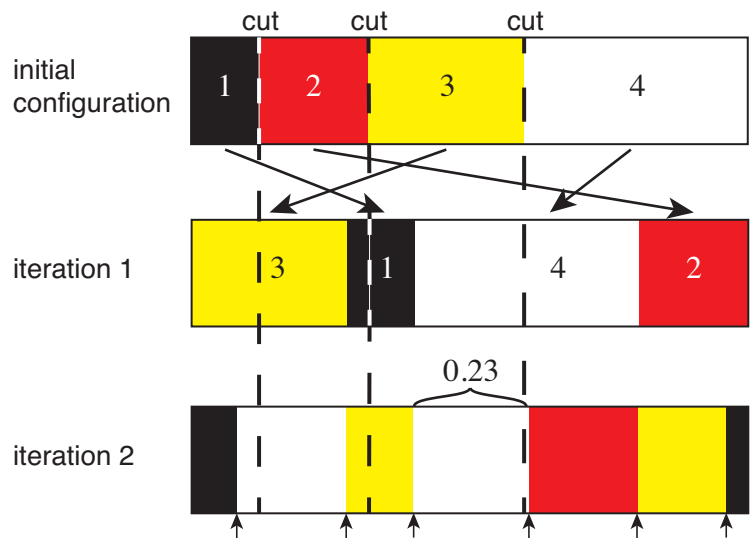}
\end{center}
\caption{(Color online.) Illustration of how the first two iterations of our IET construction with $N = 4$, $r = 1.5$ and $\Pi([1234])=[3142]$ proceed. After the second iteration, the maximum percent unmixed (corresponding to the subsegment delineated by the curly brace in the bottom row) is $U_2 = 23\%$, and number the number of interfaces (denoted by arrows at the bottom) is $C_2 = 6$.}
\label{mk:fig:cands_ex}
\end{figure}

To visualize the process of cutting and shuffling, the initial subsegments (after $\mathcal{I}$ is first subdivided into the set $S$) are each assigned a different color, as shown in Fig.~\ref{mk:fig:ic}. One application of the IET $T_{S,\Pi}$ shuffles the subsegments according to the prescribed rearrangement order given by the permutation $\Pi$.  The subsegments after the initial subdivision of $\mathcal{I}$ are consecutively numbered from left to right, as shown in the top row of Fig.~\ref{mk:fig:cands_ex}. Hence, the rearrangement order defines a one-to-one mapping from the set $\{1,2,\hdots,N\}$ into itself. The image of $\{1,2,\hdots,N\}$ under $\Pi$ is the order of the subsegments after one iteration of the cutting and shuffling process, i.e., after one application of the IET. For example, in Fig.~\ref{mk:fig:cands_ex}, the rearrangement order is given by $\Pi([1234])=[3142]$, where the bracket notation (rather than set notation) is used for the permutation to emphasize that the order of both the input and output integers is important and fixed. 

After the initial subdivision and shuffling of $\mathcal{I}$, the line segment is reassembled, then it is cut once again at the same locations, and the process continues. The result of a second application of the IET is shown in the bottom row of Fig.~\ref{mk:fig:cands_ex}. The cutting and shuffling process  continues until either a certain number of iterations have been completed, or the line segment is ``mixed'' to a predetermined level, pending a proper definition of ``mixed.'' Details of the numerical procedure used to simulate cutting and shuffling of a line segment are provided in Appendix~A.

To quantify the degree of mixing after $n$ iterations, we use two measures: the percent unmixed $U_n$ and number of distinct cuts (i.e., interfaces) $C_n$. The percent unmixed is the longest continuous subsegment of like color, even if that subsegment comes about as a result of two like-color subsegments having been recombined. Since the length of $\mathcal{I}$ was normalized to 1, the length of longest piece of material of like color corresponds to the largest percent of the material that is ``unmixed.'' Ideally, $U_n \to 0$ as $n\to\infty$ would imply complete mixing. For example, in Fig.~\ref{mk:fig:cands_ex}, the central unmixed portion after $n=2$ iterations is the longer white interval, giving $U_2 = 23\%$. 

The number of interfaces $C_n$, on the other hand, is the number of locations where subsegments of different initial color are in contact.  Locations where pieces of subsegments of like color reassemble are not included in the calculation of $C_n$, even though a cut was formally made there at some previous iteration. In contrast to $U_n$, a larger value of $C_n$ corresponds to better mixing because many pieces of differing color are in contact. In Fig.~\ref{mk:fig:cands_ex}, after 2 iterations, we have $C_2 = 6$.

A number of alternative mathematical definitions of the quality of mixing can found in the literature (see, e.g., \cite[\S1.2]{ank02}). Recently, mathematically-rigorous measures of mixing capable of quantifying the ``multiscale'' nature of the self-similar lamellar structures developed by stretching and folding in chaotic flows \cite{mmso92} have been developed using the tools of dynamical systems \cite{mmp05} and partial differential equations \cite{jlt12}. For our purposes, however, $U_n$ and $C_n$ are easy to compute, require few assumptions, are conceptually simple, and have concrete physical meaning. Thus, we use them to quantify the complexity generated by cutting and shuffling a line segment.


\section{Summary of mathematical results on IETs}
\label{mk:sec:theory}

In this section, we review the pertinent previous mathematical results that characterise IETs to aid in the interpretation of our computational results. Specifically, we introduce the necessary terminology and the key theorems applicable to our study.

An in-depth discussion of the \emph{ergodic hierarchy} that defines ``levels'' of mixing using mathematically-precise concepts from dynamical systems theory can be found in \cite[\S3.7]{sow06} and \cite{bfk06}. An elementary discussion in relation to IETs and PWIs can be found in \cite[\S V]{clo11}. Keane's minimality condition \cite[\S2]{k75} is an important result in this respect as it guarantees (unique) ergodicity, the ``weakest'' form of mixing, of (almost all) IETs \cite{m82,v82}. The ``almost all'' part the latter result is strict as non-trivial IETs that are minimal but not (uniquely) ergodic have been constructed \cite{k77}. We give a short summary of the background on these results through some equivalent and easy-to-state theorems due to \citet{v06}, after providing some necessary definitions.

\begin{definition}
A permutation $\Pi$ is said to be \emph{irreducible} if applying $\Pi$ to any of the subsets $\{1\}$, $\{1,2\}$, $\{1,2,3\}$ up to $\{1,2,\hdots,N-1\}$ does not yield a permutation of just the elements of \emph{that} subset.
\label{mk:def:irred}
\end{definition}

By $\Pi(i)$ we denote the value that the integer $i$ is mapped to by the permutation, while the compacted notation $\Pi([123\hdots])$ is used to show (all at once) the image under the action of $\Pi$ of the integers $\{1,2,3,\hdots\}$. For example, the permutation $\Pi([12345]) = [31254]$ is reducible (i.e., \emph{not irreducible}) because the first three elements (and the last two elements) are a permutation of only themselves (neither 1, 2, nor 3 maps to 4 or 5). In contrast, $\Pi([12345]) = [31524]$ is \emph{irreducible}. See Appendix~B for further details. 

\begin{definition}
A permutation $\Pi$ of $\{1,2,\hdots,N\}$ is said to be a rotation if $\Pi(i+1) \equiv \Pi(i)+1 \mod N$.
\label{mk:def:rot}
\end{definition}

For example, it easy to show that $\Pi([12345]) = [34512]$ satisfies this definition, hence it is a rotation. On the other hand, $\Pi([12345]) = [54312]$ violates the definition for $i=1$ because $2$ maps to $\Pi(2) = 4$ while $1$ maps to $\Pi(1) = 5$ and $4\ne 5 + 1 \mod 5 = 1$, hence $\Pi([12345]) = [54312]$ is not a rotation.

\begin{theorem}[\citealp*{v06}, Proposition 3.2]
If the lengths of the subsegments $\mathcal{I}_i\in S$ are rationally independent and $\Pi$ is irreducible, then $S$ and $\Pi$ satisfy the \emph{Keane minimality condition}.
\label{mk:thm:keane_cond_def}
\end{theorem}

For example, to ensure that the lengths of subsegments $\mathcal{I}_i\in S$ are rationally independent, we could require that the lengths of adjacent intervals $r_i=|\mathcal{I}_{i}|/|\mathcal{I}_{i-1}|$ be irrational numbers. Here, we will not state Keane's condition in its original form \cite[\S2]{k75} because Theorem~\ref{mk:thm:keane_cond_def} is more accessible. Furthermore, Theorems \ref{mk:th:minimal} and \ref{mk:th:noperiod} below characterize all IETs satisfying Keane's condition, which allows us to bypass its cumbersome mathematical definition.

\begin{theorem}[\citealp*{v06}, Proposition 4.1]
If $S$ and $\Pi$ satisfy the Keane condition, then every orbit of $T_{S,\Pi}$ is dense in the whole domain; such an IET is termed \emph{minimal}.
\label{mk:th:minimal}
\end{theorem}

Here, by ``orbit'' we mean the set of points $\{y,T_{S,\Pi}(y),T_{S,\Pi}^2(y),\hdots,T_{S,\Pi}^n(y),\dots\}$ for a given $y\in\mathcal{I}$. An orbit is dense in the domain if, for some $n$, it comes arbitrarily close to any given point in the domain.

\begin{theorem}[\citealp*{v06}, Lemma 4.4]
If $S$ and $\Pi$ satisfy the Keane condition, then $T_{S,\Pi}$ has no periodic points.
\label{mk:th:noperiod}
\end{theorem}

Thus, if an IET satisfies the conditions of Theorem \ref{mk:thm:keane_cond_def}, it satisfies the Keane condition, meaning (by Theorems \ref{mk:th:minimal} and \ref{mk:th:noperiod}) that ``all its orbits visit all of the domain'' and, specifically, none are periodic. This would suggest good mixing because material is distributed throughout all of the interval. However, we could be more precise if we consider the map's behavior in the limit as the number of iterations goes to infinity. Let $\mathcal{A}_{1},\mathcal{A}_{2}\subset\mathcal{I}$ be \emph{any} two subintervals of $\mathcal{I}$, where $\mathcal{I}=[0,1]$ in our construction. (These subintervals are arbitrary and do not have to coincide with the subsegments $\mathcal{I}_i$.) Denoting the cutting and shuffling process (i.e., the IET $T_{S,\Pi}$) simply by $T$, the condition for \emph{strong} mixing can be stated mathematically as
\begin{equation}
\lim_{n\to\infty} \mu\big(T^n(\mathcal{A}_1)\cap\mathcal{A}_2 \big) = \mu(\mathcal{A}_1)\mu(\mathcal{A}_2),
\label{mk:eq:math_mixing}
\end{equation}
where $\mu(\cdot)$ is the appropriate \emph{invariant measure}, e.g., length in 1D, and $T^n(\mathcal{A}_1)\cap\mathcal{A}_2$ denotes the material in common between $T^n(\mathcal{A}_1)$ and $\mathcal{A}_2$.
In other words, as any subinterval $\mathcal{A}_1$ is cut and shuffled throughout the domain $\mathcal{I}$, we would expect to find the same amount of material from $\mathcal{A}_1$ in any other subinterval $\mathcal{A}_2$.

A cutting and shuffling map based on an IET cannot be strongly mixing \cite{k80}. However, the following can be shown.
\begin{theorem}[\citealp*{af07}, Theorem A]
Let $\Pi$ be an irreducible permutation that is not a rotation. Then, for almost every $S$, $T_{S,\Pi}$ is \emph{weakly mixing}, i.e.,
\begin{equation}
\lim_{n\to\infty} \frac{1}{n}\sum_{i=0}^{n-1} \big|\mu\big(T^i(\mathcal{A}_1)\cap\mathcal{A}_2 \big) - \mu(\mathcal{A}_1)\mu(\mathcal{A}_2) \big| = 0.
\label{mk:eq:weak_math_mixing}
\end{equation}
\label{mk:th:wm}
\end{theorem}
Put simply, Theorem~\ref{mk:th:wm} relaxes the requirement that the same amount of material from the subinterval $\mathcal{A}_1$ be found in any other subinterval $\mathcal{A}_2$, to the requirement that Eq.~\eqref{mk:eq:math_mixing} hold except for a countable number of $n\in\mathbb{N}$ \citep[p.~45]{w82}. Still, weak mixing is a stronger result than the ergodicity property mentioned previously that an IET acquires by satisfying Keane's minimality condition. Again, the ``almost all'' requirement is strict as there are non-trivial minimal IETs that are (uniquely) ergodic but not weakly mixing \cite{hmli10}. The situation becomes much more complicated in 2D, where only a few results are known for \emph{rectangle exchange transformations} \cite{h81} and almost none for PWIs. However, there is at least one example, due to \citet{k02}, of a PWI that is \emph{not} weakly-mixing.

The last mathematical result we review concerns the growth of the number of interfaces $C_n$ with $n$. It is characterized by the following theorem due to \citet{n09}.
\begin{theorem}[\citealp*{n09}, Theorem 1.1]
For any interval exchange transformation $T_{S,\Pi}$, either $C_n(T)$ exhibits linear growth
in $n$ or $C_n(T)$ is bounded above independently of $n$.
\label{mk:th:growth}
\end{theorem}

In the case when $C_n$ exhibits linear growth, the slope is given by the number of \emph{nonresolving fundamental discontinuities} \cite[Proposition 2.3]{n09}, i.e., those cuts at which two like colors are never ``glued back together,'' as $n\to\infty$. Unlike chaotic dynamical systems in which any analog of $C_n$ grows exponentially with $n$, quickly leading to an intractable numerical problem in 2D or 3D \cite{fo87}, Theorem~\ref{mk:th:growth} ensures that keeping track of the interfaces generated by an IET is feasible. Finally, it is important to note that the dichotomy of the growth of $C_n$ does not correspond directly to the dichotomy of whether $T_{S,\Pi}$ satisfies Keane's condition or not. We explore the relationship between the two through the numerical examples below.


\section{Numerical results}
\label{mk:sec:num_res}

Given the simple system of an interval divided into subsegments that we constructed in Section~\ref{mk:sec:IET_construct}, we can set about to explore the relationship between the mathematical results on IETs summarized in Section~\ref{mk:sec:theory} and mixing by cutting and shuffling. For example, a natural question to ask is what happens when Keane's condition is violated? Finite-precision arithmetic ensures that, on a computer, there are no ``truly irrational'' choices for ratio  $r$ between adjacent subsegment lengths. We would like to determine how this affects the mixing properties of IETs. In this section, we investigate several examples of IETs constructed as described in Sec.~\ref{mk:sec:IET_construct} with the goal of understanding the impact of the theorems from Sec.~\ref{mk:sec:theory} on finite-time mixing by cutting and shuffling. The latter is an aspect of IETs that is not covered by the mathematical theory. To make the computational analysis easier, we restrict ourselves to a relatively small number of subsegments $N\in\{4,5,6,7\}$ and a ratio between subsegment lengths $r \in [1,2.5]$.

\begin{figure}[h]
\begin{center}
\includegraphics{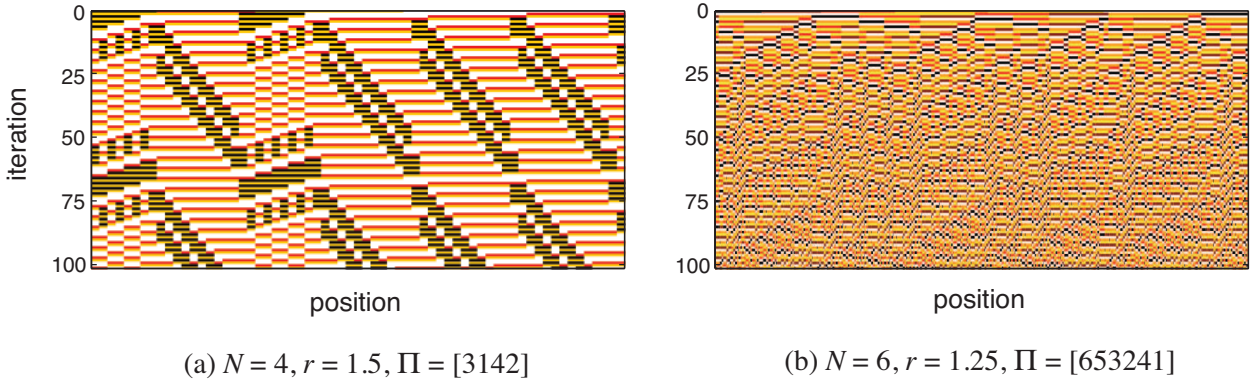}
\end{center}
\caption{(Color online.) Space-time plots over 100 iterations of our IET-based cutting and shuffling map for two different combinations of parameters. (a) Periodic dynamics and poor mixing result from choosing $N = 4$, $r = 1.5$ and $\Pi([1234])=[3142]$. (b) Substantial mixing is observed for $N = 6$, $r = 1.25$ and $\Pi([123456])=[653241]$.}
\label{mk:fig:spacetime}
\end{figure}

To visually analyze the impact of the various parameters on the mixing properties of our IET, we construct space-time plots of the cut and shuffled line segment $\mathcal{I}$ as shown in Fig.~\ref{mk:fig:spacetime}. In such space-time plots, the initial subdivision of $\mathcal{I}$ into $N$ differently-colored subsegments is at the top. The cut and shuffled versions of $\mathcal{I}$ resulting from subsequent applications of the IET, i.e., $T^n(\mathcal{I})$, are stacked below the initial condition. A small number of pieces of different colors at the bottom of a space-time plot indicates poor mixing. This is one of the most useful aspects of the space-time plot: repeating patterns in the graph correlate with poor mixing. For instance, in Fig.~\ref{mk:fig:spacetime}(a), periodic dynamics, which leads to poor mixing, are immediately evident. An IET that can produce significant mixing, on the other hand, leads to the bottom row of the space-time plot having many subsegments of different colors, as shown in Fig.~\ref{mk:fig:spacetime}(b).

It is worthwhile comparing space-time diagrams to \emph{Poincar\'e sections}, a common diagnostic of mixing. A Poincar\'e section consists of the collection of locations of ``tracers'' after each application of a map representing the action of, say, one period of a flow. When each tracer is assigned a different color, the Poincar\'e section takes a particularly visually appealing form (see, e.g., \cite[Chap.~6]{o89}). In this one-dimensional context, however, the ``tracers'' can be thought of as points distributed along a line segment, filling it densely in the case of good mixing, or leaving gaps (depending on their seed locations) in the case of poor mixing. Constructing a space-time diagram as in Fig.~\ref{mk:fig:spacetime} is a clearer way of visualizing the same information: any row represents a mixing pattern that can be compared to a previous or later iteration showing where colors, which correspond to different initial locations, are distributed under the action of cutting and shuffling.

\subsection{Effect of rearrangement order}

\begin{figure}[h]
\begin{center}
\includegraphics{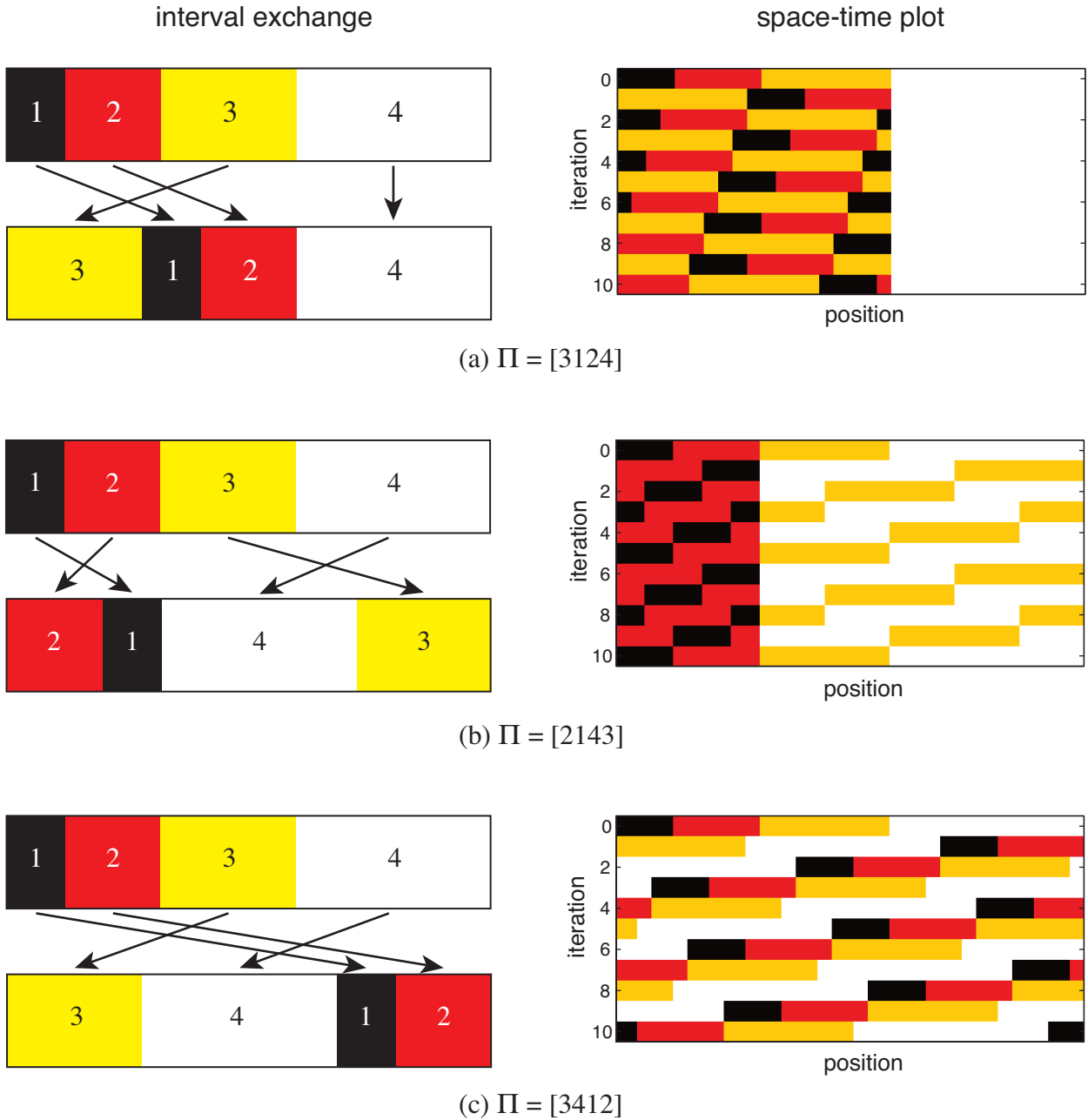}
\end{center}
\caption{(Color online.) Cutting and shuffling of a line segment with rearrangement order given by reducible permutations: (a) $\Pi([1234])=[3124]$ for which the last subsegment remains in place, (b) $\Pi([1234])=[2143]$ in which the first and last pair of subsegments are exchanged pairwise and (c) $\Pi([1234])=[3412]$ which is a rotation. In all three cases $N = 4$ and $r = 1.5$. On the right are space-time plots of the first 10 iterations of the cutting and shuffling process.}
\label{mk:fig:effectorder}
\end{figure}

The choice of the subsegment rearrangement order, represented by the permutation $\Pi$, obviously affects the quality of mixing as made precise by the theorems in Sec.~\ref{mk:sec:theory}. Permutations that are reducible, as described in Definition~\ref{mk:def:irred}, result in poor mixing and, often, periodic dynamics. Figure~\ref{mk:fig:effectorder} illustrates this behavior for three choices of $\Pi$ with $N$ and $r$ fixed. 

The simplest reducible permutations are those for which either $\Pi(1) = 1$ (meaning the first subsegment remains in place throughout the cutting and shuffling process) or $\Pi(N) = N$ (meaning the last subsegment remains in place, as shown in Fig.~\ref{mk:fig:effectorder}(a)). In either case, the percent unmixed $U_n$ is equal (or becomes equal at some $n>1$) to the length of this fixed subsegment; no further reduction in $U_n$ can be achieved. Figure~\ref{mk:fig:effectorder}(b), on the other hand, displays a different type of reducible rearrangement order in which subsegments 1, 2 and 3, 4 are exchanged pairwise. Again, $U_n$ does not tend to zero during the cutting and shuffling process. Additionally, the number of interfaces $C_n$ is easily seen to remain between 3 and 5. This is, of course, indicative of poor mixing. Permutations that are rotations, as described in Definition~\ref{mk:def:rot}, also result in poor mixing, though they do not explicitly violate Keane's condition (our choice of $r$, however, does). Figure~\ref{mk:fig:spacetime}(c) shows such a case, with the space-time plot clearly illustrating why this $\Pi$ is termed a ``rotation.'' In each iteration the last two subsegments are shifted to the beginning of the line segment, resulting in periodic dynamics and poor mixing.

\begin{figure}[b]
\begin{center}
\includegraphics{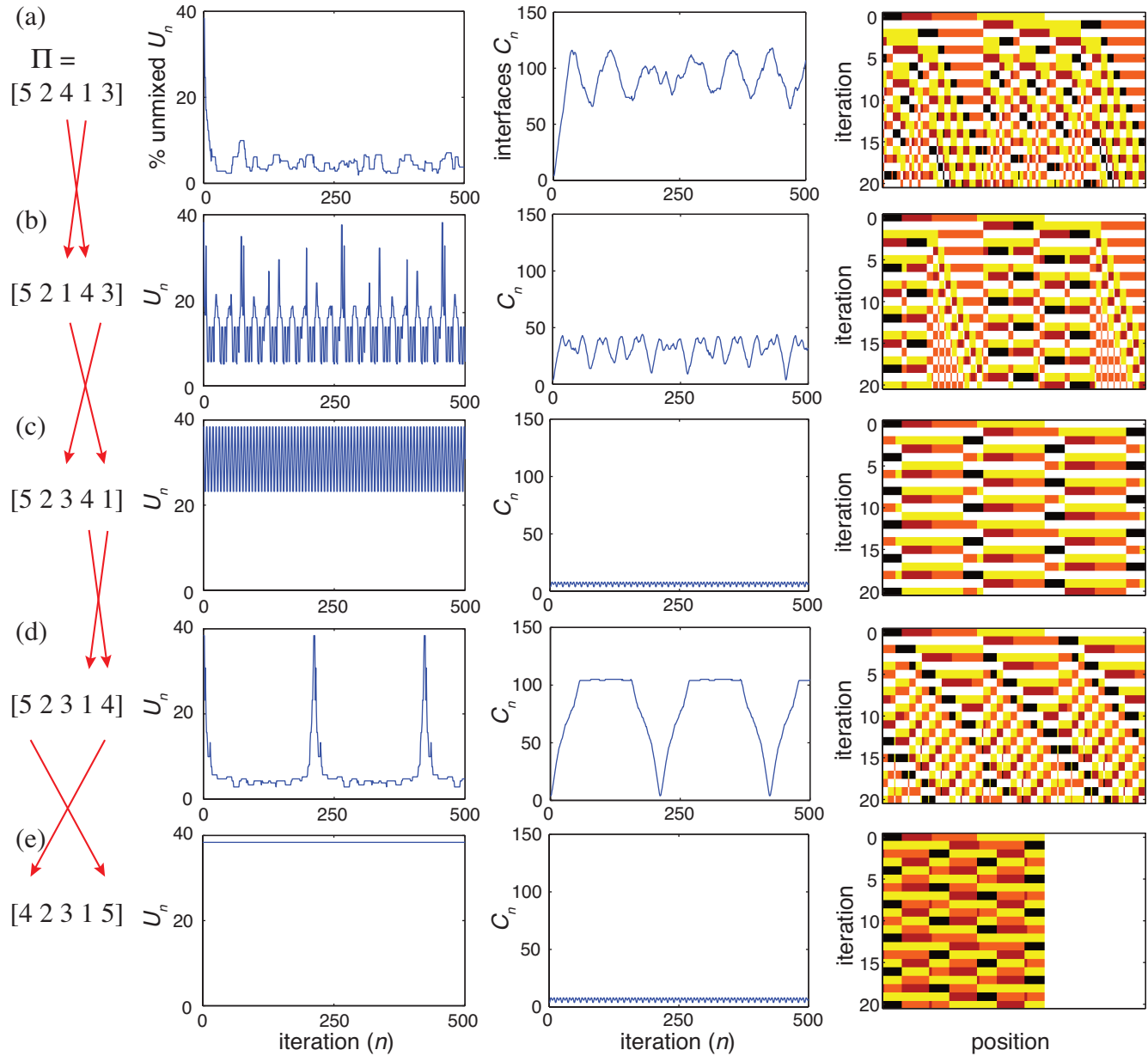}
\end{center}
\caption{(Color online.) When Keane's condition is violated by choosing $r = 1.5$ for $N = 5$, switching the locations of only 2 elements in the rearrangement order can have a significant impact on the mixing dynamics. There are five typical outcomes. (a) Significant mixing for $\Pi([12345])=[52413]$.  (b) $U_n$ and $C_n$ oscillate under the rearrangement order = $\Pi([1234])=[52143]$.  (c) Periodic dynamics result for $\Pi([1234])=[52341]$ with $U_n$ and $C_n$ oscillating at a high frequency.  (d) Significant mixing can be achieved but the dynamics are ultimately periodic (jumps in $U_n$, dips in $C_n$ always occur after a fixed number of iterations) for $\Pi([12345])=[52314]$.  (e) $U_n$ remains constant and $C_n$ is bounded above by a small number under the reducible rearrangement order $\Pi([12345])=[42315]$.}
\label{mk:fig:effectorder2}
\end{figure}

So far, as expected, we have shown that choices of $\Pi$ and $r$ that violate Keane's condition lead to trivial dynamics and poor mixing. It is worthwhile to also consider cases when the permutation is \emph{irreducible} (as required by Theorem~\ref{mk:thm:keane_cond_def}) but $r$ is still chosen to violate Keane's condition. Now, as shown in Fig.~\ref{mk:fig:effectorder2}, we find that a slight change in the rearrangement order can have a significant impact on the quality of mixing by cutting and shuffling. 

Having performed a number of numerical experiments, we find that five typical outcomes can be observed.  First, as shown in Fig.~\ref{mk:fig:effectorder2}(a), there can be significant mixing.  The percent unmixed $U_n$ decays to a small value quickly, and the number of interfaces $C_n$ grows to a large value. However, there is a clear bound (lower for $U_n$, upper for $C_n$) on the quality of mixing that can be achieved because $r = 3/2$ is a rational number. By switching the location of only two subsegments in the rearrangement order, we observe a second kind of mixing behavior, shown in Fig.~\ref{mk:fig:effectorder2}(b). Now, the quality of mixing remains poor with large-amplitude oscillations present in $U_n$ and $C_n$. This occurs because subsegments that are cut and shuffled can become reassembled at a later iteration.  Upon switching another two elements in the rearrangement order, a third type of behavior is observed in Fig.~\ref{mk:fig:effectorder2}(c): $U_n$ and $C_n$ now oscillate at a much higher frequency; $U_n$ remains high and $C_n$ remains low, indicating poor mixing.  Again, this results because the subsegments are reassembling into their initial configuration every few iterations due to the periodic dynamics of this IET. The fourth type of behavior, produced by another small change in the rearrangement order, results in the dynamics shown in Fig.~\ref{mk:fig:effectorder2}(d), where $U_n$ and $C_n$ now exhibit two distinct time scales of oscillation. One is at high frequency and with low amplitude (``local'' changes every few iterations), the other at low frequency and with large amplitude (``global'' reassemblies of the initial condition). At the iteration corresponding to a peak in $U_n$, the interval has reassembled into its original state. Though there is, apparently, significant mixing at certain $n$, periodic dynamics still exist because $r=3/2$ is a rational number, which violates  Keane's condition. Finally, Fig.~\ref{mk:fig:effectorder2}(e) presents the fifth type of behavior observed. In this case, the rearrangement order is a reducible permutation with $\Pi(N) = N$. As expected, there is very poor mixing; $U_n$ remains constant (because the last, and longest, subsegment is never cut and shuffled) and $C_n$ is bounded above by a small integer for all $n$.

\subsection{Effect of interval length ratio}

\begin{figure}
\begin{center}
\includegraphics{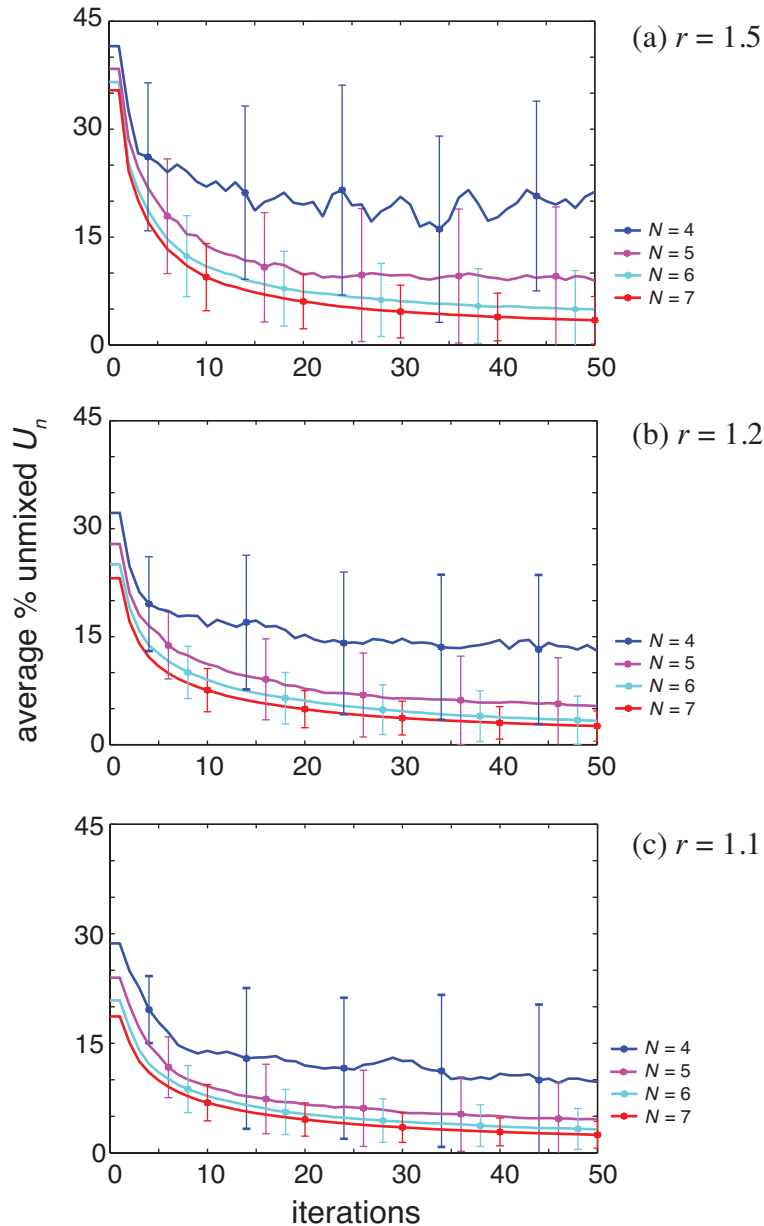}
\end{center}
\caption{(Color online.) Average (across all irreducible rearrangement orders) percent unmixed $U_n$ for (a) $r = 1.5$ and $N = 4$ to $7$, (b) $r = 1.2$ and $N = 4$ to $7$, (c) $r = 1.1$ and $N = 4$ to $6$. Error bars represent one standard deviation about the mean. $N=6$ appears to be a point of diminishing returns, and $r$ closer to 1 (a more uniform distribution of initial subsegment lengths) leads to better mixing on average.}
\label{mk:fig:effectr}
\end{figure}

Both the interval length ratio $r$ and the number of subsegments $N$ significantly affect the quality of mixing by cutting and shuffling. In Fig.~\ref{mk:fig:effectr}, the percent unmixed $U_n$ (averaged over all irreducible permutations for a given $N$) is plotted as a function of the number of iterations $n$ for several choices of $r$ and $N$.  As expected for such irreducible permutations, $U_n$ decreases with $n$; the greatest decrease occurs within the first $20$ or so  iterations. For larger $N$, the curves in Fig.~\ref{mk:fig:effectr} are systematically lower. This indicates that starting with more subsegments leads to faster mixing. Furthermore, the change in $U_n$ resulting from adding an additional segment is greater when going from $N = 4$ to $N = 5$ than when going from $N = 5$ to $N = 6$ or from $N=6$ to $N=7$.  This illustrates that there is a point of diminishing returns. The errors bars are larger for smaller $N$ because there are more protocols that mix poorly in those cases. Thus, in practice, at least 6 or 7 subsegments are necessary to obtain significant mixing across a range of protocols, but starting with more subsegments than that has minimal impact on the overall mixing. With a larger number of subsegments, the precise rearrangement order chosen in the protocol has less impact than for smaller $N$, as long as the permutation is irreducible.  

\begin{figure}
\begin{center}
\includegraphics{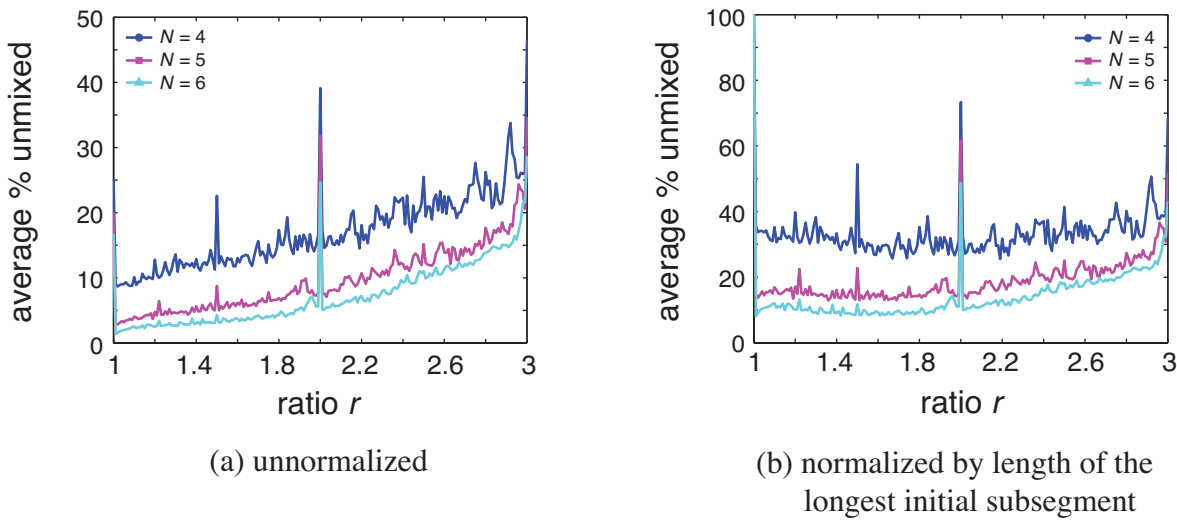}
\end{center}
\caption{(Color online.) Average (across iterations 50 to 100 \emph{and} across all irreducible rearrangement orders) percent unmixed $U_n$ versus the subsegment ratio $r$ for $N = 4$, $5$ and $6$. Arnold tongue-like phenomena are seen at the resonances $2/1$, $3/1$, $3/2$, $5/2$, etc. The best mixing (on average) occurs for $r$ close, but not equal, to $1$ because the distribution of lengths of the initial subsegments is close to uniform. The fact that the length of the longest initial subsegment increases with $r$, meaning that (on average) more iterations are needed to reach the same percent unmixed (upward trend in (a)), can be accounted by normalizing with $U_{n,\text{max}}$, as shown in (b).}
\label{mk:fig:varyr}
\end{figure}

It is also clear from Fig.~\ref{mk:fig:effectr} that reducing $r$ results in better mixing.  To further analyze the impact of $r$, we plot $U_n$---averaged over all  irreducible rearrangement orders for a given $N$ \emph{and} then averaged from iteration 50 to 100---as a function of $r$ in Fig.~\ref{mk:fig:varyr}(a). Clearly this average $U_n$ increases as $r$ increases, with a larger number of subsegments effecting better mixing. The upward trend in Fig.~\ref{mk:fig:varyr}(a) can be understood by recalling that larger $r$ leads to an increase in the length difference between adjacent initial subsegments. In particular, it is easy to show that the longest initial subsegment, which takes the most ``effort'' to cut and shuffle, has length
\begin{equation}
U_{n,\mathrm{max}}(r,N) = r^{N-1}\left(\frac{r-1}{r^N-1}\right).
\label{mk:eq:unmax}
\end{equation}
We use the notation $U_{n,\mathrm{max}}$ because, by construction, the length of the longest initial subsegment of a given color is the upper bound on the percent unmixed for all $n$. Note that $U_{n,\mathrm{max}}$ is well behaved at $r=1$, specifically $\lim_{r\to1} U_{n,\mathrm{max}}(r,N) = 1/N$. Thus, for a fixed number of iterations (as in Fig.~\ref{mk:fig:varyr}, where all averages are calculated from iteration 50 to 100), increasing $r$ should increase the average percent unmixed. To account for this effect, in Fig.~\ref{mk:fig:varyr}(b), we divide the average percent unmixed by $U_{n,\mathrm{max}}(r,N)$. Clearly, this normalization eliminates part of the upward trend. It is natural to interpret the normalized average $U_n$ in Fig.~\ref{mk:fig:varyr}(b) as a \emph{mixing efficiency}, i.e., a measure of the average shortest segment length produced by cutting and shuffling compared to the initial longest segment length. On the other hand, from a practical standpoint, we want the unmixed portion to be as short as possible, leading to a preference for $r$ being close, but not exactly equal, to one, as indicated in Fig.~\ref{mk:fig:varyr}(a) and discussed shortly.

In addition, it is evident that certain ratios, e.g., $r=3/2$, $2/1$, $5/2$ or $3/1$, lead to particularly poor mixing, resulting in peaks in Fig.~\ref{mk:fig:varyr}, similar to the resonances leading to \emph{Arnold tongues} in a chaotic dynamical system \cite{pr07}. This corroborates the theoretical results from Sec.~\ref{mk:sec:theory} that in order to get the best mixing possible (specifically, to eliminate periodic orbits and ensure ergodicity), $r$ should be an irrational number. Furthermore, for the present IET construction, Fig.~\ref{mk:fig:varyr}(a) suggests that an irrational number close to unity is preferable. The reason for this is the manner in which we define $r$. For $r$ close to unity, the initial distribution of subsegment lengths is almost uniform, meaning no subsegment is very long compared to any other. The opposite case case ($r \gg 1$) is undesirable because long subsegments can take many iterations to be cut and shuffled. After normalizing by $U_{n,\mathrm{max}}$ in Fig.~\ref{mk:fig:varyr}(b), we observe that the mixing efficiency is still better for ratios closer to one. However, there is little difference in the normalized unmixed portion for $r\in[1,2]$. It is quite evident that as $r$ increases beyond 2, the second longest subsegment is only a small fraction of the longest subsegment. Hence, it takes many more iterations to significantly reduce the length of the longest subsegment. In the opposite extreme, as $r\to1$, the difference in length between the longest subsegment and the shortest subsegment becomes small, which makes it possible to accomplish good mixing after an adequate number of iterations. Overall, we can conclude that the mixing quality and efficiency are improved (on average) by starting with a more uniform distribution of subsegment lengths.

\subsection{Growth of the number of interfaces}
\label{mk:sec:growthcuts}

As discussed earlier, an important aspect of the mixing properties of a cutting and shuffling process is the number of interfaces it generates between materials of different color. Such interfaces arise from the discontinuities in the cutting and shuffling map, specifically where the cuts are placed in the interval exchange transformation. For PWIs, the higher-dimensional analogue of IETs that arise in the context of granular mixing in a spherical container, it was shown that the discontinuities in the map are the sole source of intermaterial interfaces in the absence of the ``usual'' chaotic dynamics \cite[\S8.9]{c11}. The recent work by \citet{n09} on the growth of the number of discontinuities as $n\to\infty$, which we summarized in Sec.~\ref{mk:sec:theory}, is the only mathematical characterization of the set of interfaces. Beyond that, determining the growth of $C_n$ is, in general, an open problem.

In this section, we aim to better understand the behavior of $C_n$ by supplementing the mathematical theorems with numerical results. Specifically, we wish to determine what effect violating Keane's condition has on the growth of $C_n$, and whether violating it to different ``degrees'' matters. To this end, we take $N$ fixed and consider the two cases of a given reducible or a given irreducible rearrangement order $\Pi$. In each case, we vary the value of $r$. Ideally, $r$ should be irrational to observe the asymptotic results of \citet{n09}. To approach this limit, we take $r$ to be increasingly ``more irrational'' and compute the evolution of $C_n$ for the given $N$ and $\Pi$.

To make the idea of ``more irrational'' precise, we apply the concept of a \emph{finite continued fraction} of $k$ partial denominators, which is defined as 
\begin{equation}
[a_0;a_1,\hdots,a_k] := a_0 + \cfrac{1}{a_1 + \cfrac{1}{a_2 + \cfrac{1}{ \ddots + \cfrac{1}{a_k} }}}.
\end{equation}
The continued fraction is \emph{simple} if $\{a_0,\hdots,a_k\}$ are all integers. Every real number has a unique simple continued fraction representation; rational numbers correspond to finite continued fractions, while irrational numbers correspond to infinite continued fractions \cite[Chap.~15]{b07}. We call one rational number ``more irrational'' than another if it has a longer continued fraction expansion.

\begin{figure}[h]
\begin{center}
\includegraphics{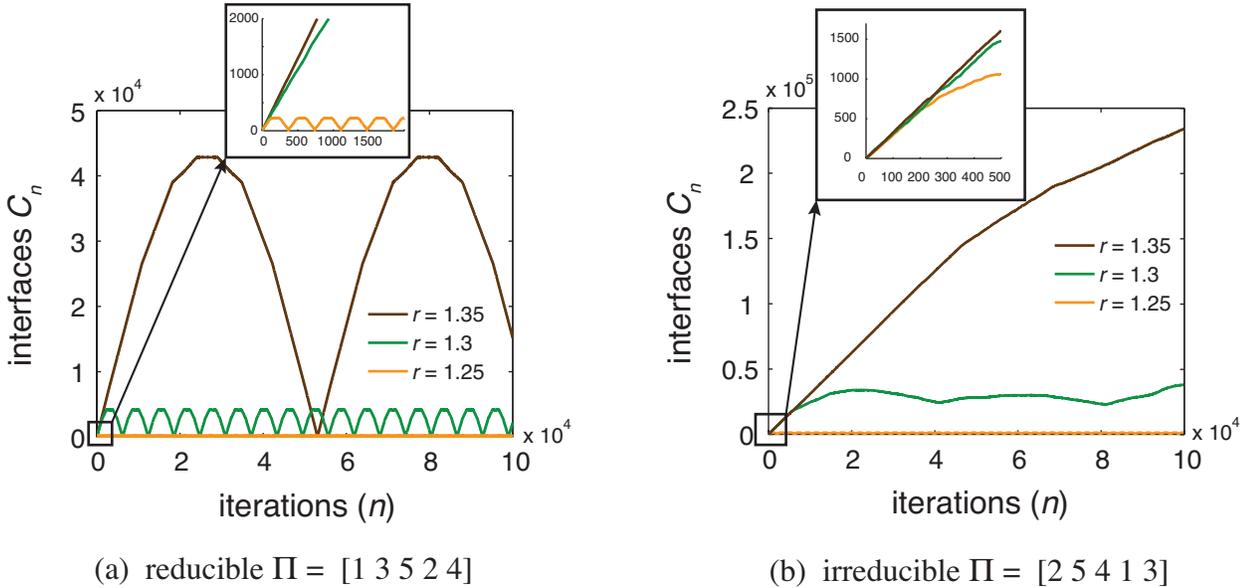}
\end{center}
\caption{(Color online.) Number of interfaces $C_n$ plotted as a function of iterations for $N=5$, ratios $r = 1.25$ (orange), $1.3$ (green), $1.35$ (brown) and (a) the reducible rearrangement order $\Pi([12345]) = [13524]$ and (b) the irreducible rearrangement order $\Pi([12345]) = [25413]$. Note that, for clarity, different vertical scales are used in panels (a) and (b).}
\label{mk:fig:growthcuts}
\end{figure}

Thus, in Fig.~\ref{mk:fig:growthcuts}, we consider the following three choices for the ratio of subsegment lengths: $r = 1.25 = 5/4 = [1;4]$ as the ``least irrational,'' $r = 1.3 = 13/10 = [1;3,3]$ as the ``more irrational'' and $r = 1.35 = 27/20 = [1;2,1,6]$ as the ``most irrational.'' Figure~\ref{mk:fig:growthcuts}(a) shows the growth of $C_n$ for the protocol generated by the reducible permutation $\Pi([12345]) = [13524]$. In this case, the dynamics is eventually periodic for all three choices of $r$ with the period of $C_n$ becoming longer as $r$ becomes ``more irrational.'' 

Meanwhile, for the irreducible permutation $\Pi([12345]) = [25413]$ in Fig.~\ref{mk:fig:growthcuts}(b), we do not observe periodic dynamics for $r=1.3$ and $r=1.35$ over the $10^5$ iterations considered. For $r=1.3$, however, $C_n$ is bounded above independently of $n$, which leads us to conjecture that this case of Theorem~\ref{mk:th:growth} corresponds to IETs that violate Keane's condition. For $r=1.35$, we expect that there is also an upper bound for $C_n$ that is independent of $n$, but it is not achieved over the $10^5$ iterations considered. Thus, this last protocol can be considered, for all practical purposes, to behave as if Keane's condition were satisfied. As a consequence, this is strong evidence that the dichotomy of grow--no growth in Theorem~\ref{mk:th:growth} corresponds to the dichotomy of satisfying--violating the assumptions of Theorem~\ref{mk:thm:keane_cond_def} (equivalently, Keane's minimality condition).

The inset in Fig.~\ref{mk:fig:growthcuts}(b) shows that the slope of all three $C_n$ curves for small $n$ is essentially independent of $r$. This suggests the interesting possibility that the slope is fixed by the number of \emph{fundamental discontinuities} (in the language of \citet{n09}), which is encoded by the chosen irreducible permutation $\Pi$ alone. All three protocols in Fig.~\ref{mk:fig:growthcuts}(b) violate Keane's condition to various ``degrees,'' and thus the number of fundamental discontinuities is, technically speaking, zero because these protocols must be eventually periodic. Indeed, the linear growth of $C_n$ is sustained over a shorter range of $n$ as $r$ is made ``more rational.'' Nevertheless, we observe a similar slope initially, which suggests that if $r$ were irrational, then this is the growth rate of $C_n$ and is thus the number of fundamental (non-resolvable) discontinuities permitted for this choice of $N$ and $\Pi$.

\subsection{Statistics on finite-time mixing}
\label{mk:sec:finitetime}

As mentioned in the Introduction, the mathematical theory (Sec.~\ref{mk:sec:theory}) does not provide an answer to questions such as ``How many iterations does it take to reduce the longest continuous-color segment in an IET to a given fraction of the interval?''. In this section, we address this question through numerical simulations. Following \cite{jcol12}, we evaluate the quality of mixing of \emph{all} the cutting and shuffling protocols that can be constructed for given $N$ and $r$ by plotting the percentage of protocols (i.e., percentage of permutations $\Pi$ for fixed $N$ and $r$) that achieve a given percent unmixed value $U_n$ after $n=50$ iterations. 

\begin{figure}[h]
\begin{center}
\includegraphics{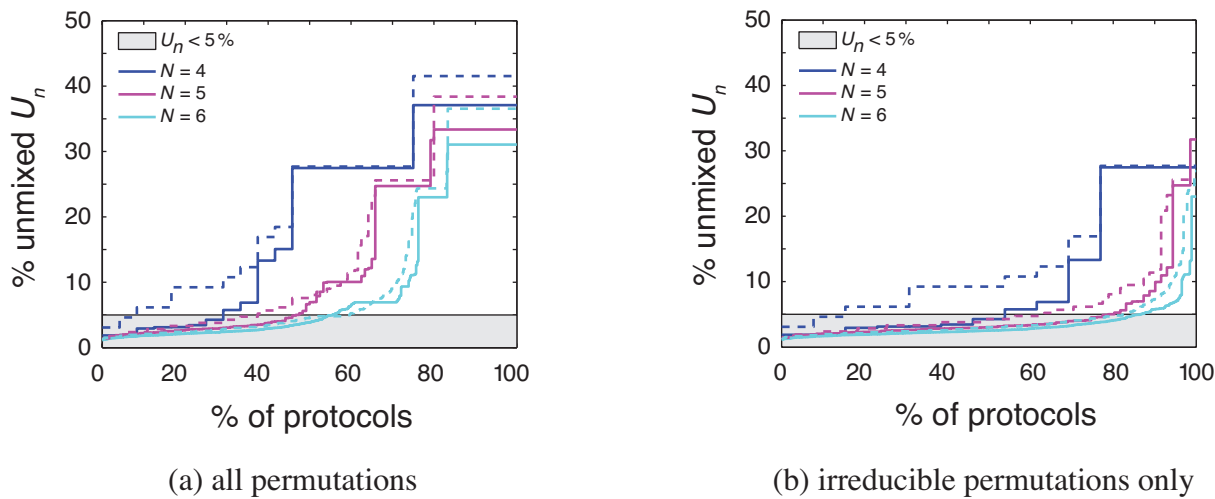}
\end{center}
\caption{(Color online.) Statistics on finite-time mixing by cutting and shuffling showing the percent unmixed achieved by a given percentage of protocols after $50$ iterations, where the ratio $r$ and number of initial subsegments $N$ is given and (a) all permutations or (b) only irreducible permutations are considered for the rearrangement order. Solid and dashed curves correspond to $r = 1.35$ and $r=1.5$, respectively.}
\label{mk:fig:protunmixed}
\end{figure}

Figure~\ref{mk:fig:protunmixed} shows the finite-time mixing statistics for the protocols with $r=1.5$ (a ``bad'' choice, corresponding to a local maximum in Fig.~\ref{mk:fig:varyr}) and $r=1.35$ (a ``better'' choice) with $N=4$, $5$ and $6$. Both of these ratios are rational numbers, and we have thus violated the assumptions of Theorems~\ref{mk:thm:keane_cond_def} and \ref{mk:th:wm}. Consequently, these IETs are neither ergodic nor weakly mixing in the mathematical sense. Nevertheless, the majority of these protocols achieve a substantial amount of mixing (as quantified by $U_n$) over 50 iterations. Furthermore, there is a clear benefit to picking $r$ to be ``more irrational'' (as discussed in Sec.~\ref{mk:sec:growthcuts}) because the curves for $r=1.35$ (solid) consistently lie below those for $r=1.5$ (dashed), indicating a higher degree of mixing across all protocols after 50 iterations.

The ``worst case scenario'' corresponds to the intersection of a curve with the vertical line at 100\% of protocols. No protocols can mix any less than the percent of the domain corresponding to the longest initial subsegment. Based on the IET construction from Sec.~\ref{mk:sec:IET_construct}, we know that the length of longest subsegment at $n=0$ is $U_n = U_{n,\mathrm{max}}$ as defined in Eq.~\eqref{mk:eq:unmax}. Thus, 100\% of protocols must achieve \emph{at least} this value of $U_n$. From Eq.~\eqref{mk:eq:unmax} we find that $U_{n,\mathrm{max}}(r=1.35,N=4,5,6) \approx 37, 33, 31$\% and $U_{n,\mathrm{max}}(r=1.5,N=4,5,6) \approx 41.5,38,36.5$\%. Indeed, these are the highest values (i.e., $U_n$ at 100\% of protocols) observed in Fig.~\ref{mk:fig:protunmixed}(a). However, it is clear that most protocols do much better, and the percentage of protocols that effect significant mixing, say those falling below the $5\%$ unmixed line in the shaded area, is much higher when only irreducible permutations are considered in Fig.~\ref{mk:fig:protunmixed}(b). No protocols achieve complete mixing over the 50 iterations considered, therefore all the curves have a non-zero limiting value on the left. However, it is evident that in all cases there exists at least one protocol that achieves $U_n < 5$\% because limiting value as \% of protocols $\to0$ is always in the shaded portion of the graph.

This last observation allows us to determine bounds on the \emph{95\% mixing time}, which is defined as the first iteration $n^*$ for which $U_{n^*} < 5\%$ \cite{ank02}. The shaded portions of Fig.~\ref{mk:fig:protunmixed} correspond to $U_n < 5\%$, and it is evident that for $N=5$ and 6 (and both $r=1.5$ and $1.35$), the majority (i.e., over 50\%) of \emph{all} protocols achieve this value of $U_n$ in at most $50$ iterations. Therefore, the 95\% mixing time is, on average, $\lesssim 50$ iterations. When only irreducible permutations are considered for the rearrangement order (Fig.~\ref{mk:fig:protunmixed}(b)), we see that nearly 80\% of the protocols for $N=5$ and 6 (and both $r=1.5$ and $1.35$) reduce $U_n$ below 5\% over 50 iterations. Consequently, if a cutting and shuffling map can achieve the 95\% mixing time, then, in practice, we expect the presence of diffusion (or other kind of irreversibility) to result in complete and thorough mixing, even if the cutting and shuffling protocol may be eventually periodic (see, e.g., the relevant discussion and examples in \cite{ank02,sturman12}).


\section{Conclusion}
\label{mk:sec:conclusion}

Although the problem of mixing a line segment by cutting and shuffling seems simple at a first glance, this first comprehensive numerical study of cutting and shuffling protocols based on interval exchange transformations (IETs) demonstrates a breadth of possible dynamical behaviors. Some of these behaviors can be predicted by the mathematical theory of IETs, while others could only be found through numerical experimentation. In addition to observing and classifying the dynamics of this cutting and shuffling process, we quantified its finite-time mixing properties within a well-defined parameter space. This is not only of practical interest in determining how cutting and shuffling leads to complex dynamics but is also of abstract interest because it could stimulate new theoretical results in the area of IETs.

Based on our study, the following design rules for mixing a line segment by cutting and shuffling can be drawn:
\begin{romanlist}[(iii)]
    \item The rearrangement order should be an irreducible permutation (see Appendix~B for more technical details on this topic).
	\item An IET with $N=6$ subsegments is the most practical number of subsegments. The improvement in mixing diminishes with larger $N$.
	\item The ratio of adjacent subsegment lengths should not be large, specifically it should be an irrational number slightly larger than 1. (Continued fraction expansions can be used to systematically make $r$ ``more irrational.'')
\end{romanlist}

Thus, an important result of the present work is that even if a cutting and shuffling map cannot be proven to be mixing in the mathematical sense, it can generate significant complexity over a finite number of iterations. Specifically, for all practical purposes it can behave (over a finite number of iterations) as if satisfied the mathematical requirements for mixing (see Sec.~\ref{mk:sec:growthcuts}). Thus, in physical systems where cutting and shuffling is one of several underlying mixing mechanisms (e.g., in granular flows \cite{col10,col10b,jlosw10,jcol12} as motivated in the Introduction), optimizing for cutting and shuffling is beneficial and, indeed, desirable. If the cutting and shuffling process is in some sense ``optimal,'' then the additional mixing mechanisms that become relevant at later times can achieve greater efficiency. The design rules listed above can have an impact in performing such optimization of physical mixing processes, once the underlying cutting and shuffling skeleton is uncovered.

At the same time, the present study suggests several mathematical questions regarding IETs that have not been answered in the pure mathematics literature:
\begin{romanlist}[(iii)]
\item Is there a rigorous connection between resonances in chaotic dynamical systems and the spikes of the average of $U_n$ in Fig.~\ref{mk:fig:varyr} at values of $r$ that are simple fractions?
\item Is the number of nonresolving fundamental discontinuities of an IET solely a function of $\Pi$? If so, what kind?
\item Are there conditions that can be imposed on subsegment length ratios $r_i$ and the permutation $\Pi$ so that $U_n$ is less than a given percentage after $n$ iterations?
\end{romanlist}

Finally, a possible avenue of future research is to extend the present methodology to study mixing by cutting and shuffling of the unit square $\mathcal{I}\times\mathcal{I}=[0,1]^2$. This could be accomplished by, for example, applying an interval exchange transformation in the $y$-direction and extending it in the $x$-direction by making each subinterval $\mathcal{I}_i$ into a rectangle of unit horizontal length \cite{clo11}. More generally, such an interval exchange in the $y$-direction can be sequentially composed with another in the $x$-direction to produce a special case of the more general class of \emph{rectangle exchange transformations} (RETs) \cite{h81}. In comparison to IETs, for which there are a number of definitive theoretical results (recall Sec.~\ref{mk:sec:theory}), RETs are poorly understood. Thus, providing concrete numerical results could be quite useful. In addition, adding irreversibility to the cutting and shuffling process, e.g., through diffusion as in \cite{ank02,sturman12}, opens further possibilities. It is conceivable that the competition between cutting and shuffling and  diffusion would lead to \emph{one-dimensional strange eigenmodes}, which have been found to be important in understanding mixing in open chaotic flows \cite{gdtr09}.


\nonumsection{Acknowledgments} \noindent This work was supported, in part, by NSF Grant CMMI-1000469. Additionally, I.C.C.\ received partial support from NSF Grant DMS-1104047 as well. M.K.K.\ was an undergraduate Murphy Institute Scholar at the Robert R.\ McCormick School of Engineering and Applied Science at Northwestern University during the completion of this work. The authors wish to thank Paul Umbanhowar for insightful comments on the manuscript and Stephen Wiggins for remarks on various definitions of chaos. I.C.C.\ acknowledges helpful discussions with Rob Sturman on the mathematical aspects of PWIs and IETs, and thanks Jean-Luc Thiffeault for bringing to his attention the work on cutoffs in 2D chaotic fluid mixing.

\nonumsection{Appendices} 

\appendix{}
\label{app:numerical}

In this Appendix, we outline how a numerical simulation of the cutting and shuffling of a line segment by an interval exchange transformation (IET) as constructed in Sec.~\ref{mk:sec:IET_construct} can be performed. The code takes the following inputs: the adjacent subsegment ratio $r$ (a finite-precision decimal), the number of subsegments $N$ (an integer $>1$), the rearrangement order $\Pi$ (a permutation of the integers $\{1,\hdots,N\}$) and the desired number of iterations $n_\mathrm{max}$ of the process (an integer $>1$). Then, it proceeds through the steps below.

\begin{enumerate}
	\item Convert $r$ to a fraction $r_n/r_d$ and compute the \emph{least common denominator} (LCD) of the fractions $\{r^{i-1}x\}_{i=1}^N$ ($x$ is given by Eq.~\eqref{mk:eq:x}), which correspond to the lengths of the subsegments in the IET construction. The LCD can be found to be $\mathfrak{lcd}=(r_n^N-r_d^N)/(r_n-r_d)$. (Note that $r_n-r_d$ is always a factor of $r_n^N-r_d^N$ for integer $N$, so that $\mathfrak{lcd}$ is an integer.) The shortest piece that can be created by cutting and shuffling is thus 1 unit out of $\mathfrak{lcd}$.
	\item Create two vectors of length $\mathfrak{lcd}$, which we call {\tt segment} and {\tt newsegment}, to represent the line segment before and after shuffling, respectively. Accordingly, the initial subsegment divisions are located between cells $r^{i-1}\cdot x\cdot \mathfrak{lcd}$ and $r^{i-1}\cdot x\cdot \mathfrak{lcd}+1$ ($i=1,\hdots,N-1$) in the array.
	\item ``Color'' the interval to be shuffled by assigning all elements between $1$ and $x\cdot\mathfrak{lcd}$ (the first subsegment)  of the array {\tt segment} the value of $1$, all elements between $x\cdot\mathfrak{lcd}+1$ and $r\cdot x\cdot \mathfrak{lcd}$ (the second subsegment) the value of $2$, and so on.
	\item Using the rearrangement order $\Pi$, copy all elements corresponding to the first ($i=1$) subsegment portion of the array {\tt segment} into the appropriate parts of {\tt newsegment}; for example, for $\Pi(1)\ne1$,
	\begin{equation}{\tt newsegment}\Big((r^{\Pi(1)-1}\cdot x\cdot\mathfrak{lcd}+1):r^{\Pi(1)}\cdot x\cdot\mathfrak{lcd}\Big) = {\tt segment}(1:x\cdot\mathfrak{lcd}).\end{equation}
	(Here, a colon is used in the sense of the {\sc Matlab} array subscripting operator.) Repeat for all $i=1,\hdots,N$ to complete one iteration of the shuffling process.
	\item Determine the longest number of consecutive elements of {\tt newsegment} containing the same value and divide this number by the total number of elements ($\mathfrak{lcd}$) to obtain the percent unmixed $U_n$ for this (say, the $n$th) iteration.
	\item Calculate the number of adjacent elements of {\tt newsegment} that have different values to obtain the number of distinct cuts $C_n$ at this iteration.
	\item Set {\tt segment} = {\tt newsegment} and repeat Steps (4) through (7) until the desired number of iterations $n_\mathrm{max}$ is reached.
\end{enumerate}


\appendix{}
\label{app:irreduce}

In this Appendix, we give some remarks on irreducible permutations. The following {\sc Mathematica} code determines whether  {\tt p} is an irreducible permutation of $\{1,2,\hdots,N\}$ (i.e., it returns 0 for false or 1 for true):
\begin{verbatim}
<< Combinatorica`
IsReducibleQ[p_] := Module[{j, ans = 0},
   For[j = 1, j <= Length[p] - 1, j++,
    If[PermutationQ[Take[p, j]], ans = 1; Break[];];
   ];
   ans
  ];
\end{verbatim}
Table~B.1 lists the number of total and irreducible permutations as a function of $N$.   \citet{klaz} gives a discussion of the integer sequence corresponding to the number of irreducible permutations of length $N$ and its properties (see also entry A003319 in the On-Line Encyclopedia of Integer Sequences \cite{OEIS}) .
\begin{table}[h]
\tbl{Number of permutations of the integers $\{1,2,\hdots,N\}$.}
{\begin{tabular}{l c c c c c c c c c c c}\\[-2pt]
\toprule
$N$ &1 &2 &3 &4 &5 &6 &7 &8 &9 &10\\[6pt]
\hline\\[-2pt]
total $=N!$ &1 &2 &6 &24 &120 &720 &$5\,040$ &$40\,320$ &$362\,880$ &$3\,628\,800$\\[1pt]
irreducible &1 &1 &3 &13 &71 &461 &$3\,447$ &$29\,093$ &$27\,3343$ &$2\,829\,325$\\[1pt]
\botrule
\end{tabular}}
\label{mk:tb:irred}
\end{table}


\bibliographystyle{ws-ijbc}
\bibliography{c&s_refs.bib}

\begin{thebibliography}{71}
\newcommand{\enquote}[1]{``#1''}
\providecommand{\natexlab}[1]{#1}
\providecommand{\url}[1]{\texttt{#1}}
\providecommand{\urlprefix}{URL }
\expandafter\ifx\csname urlstyle\endcsname\relax
  \providecommand{\doi}[1]{doi:\discretionary{}{}{}#1}\else
  \providecommand{\doi}{doi:\discretionary{}{}{}\begingroup
  \urlstyle{rm}\Url}\fi

\bibitem[{Aldous \& Diaconis(1986)}]{ad86}
Aldous, D. \& Diaconis, P. [1986] \enquote{Shuffling cards and stopping times,}
  \emph{Am. Math. Monthly} \textbf{93},  333--348,
  \urlprefix\url{http://www.jstor.org/stable/2323590}.

\bibitem[{Ashwin \& Goetz(2005)}]{ag05}
Ashwin, P. \& Goetz, A. [2005] \enquote{Invariant curves and explosion of
  periodic islands in systems of piecewise rotations,} \emph{SIAM J. Appl. Dyn.
  Syst.} \textbf{4},  437--458, \doi{10.1137/040605394}.

\bibitem[{Ashwin \emph{et~al.}(2002)Ashwin, Nicol \& Kirkby}]{ank02}
Ashwin, P., Nicol, M. \& Kirkby, N. [2002] \enquote{Acceleration of
  one-dimensional mixing by discontinuous mappings,} \emph{Physica A}
  \textbf{310},  347--363, \doi{10.1016/S0378-4371(02)00774-4}.

\bibitem[{Avila \& Forni(2007)}]{af07}
Avila, A. \& Forni, G. [2007] \enquote{Weak mixing for interval exchange
  transformations and translation flows,} \emph{Ann. Math.} \textbf{165},
  637--664, \doi{10.4007/annals.2007.165.637}.

\bibitem[{Baker \& Gollub(1996)}]{bg96}
Baker, G.~L. \& Gollub, J.~P. [1996] \emph{Chaotic Dynamics: An Introduction},
  2nd ed. (Cambridge University Press, Cambridge).

\bibitem[{Berkovitz \emph{et~al.}(2006)Berkovitz, Frigg \& Kronz}]{bfk06}
Berkovitz, J., Frigg, R. \& Kronz, F. [2006] \enquote{The ergodic hierarchy,
  randomness and {Hamiltonian} chaos,} \emph{Stud. Hist. Phil. Mod. Phys}
  \textbf{37},  661--691, \doi{10.1016/j.shpsb.2006.02.003}.

\bibitem[{Bressaud \& Poggiaspalla(2007)}]{bp07}
Bressaud, X. \& Poggiaspalla [2007] \enquote{A tentative classification of
  bijective polygonal piecewise isometries,} \emph{Exp. Math.} \textbf{16},
  77--99, \doi{10.1080/10586458.2007.10128987}.

\bibitem[{Brown \& Chua(1996)}]{cb96}
Brown, R. \& Chua, L.~O. [1996] \enquote{Clarifying chaos: Examples and
  counterexamples,} \emph{Int. J. Bifurcation Chaos} \textbf{6},  219--249,
  \doi{10.1142/S0218127496000023}.

\bibitem[{Brown \& Chua(1998)}]{cb98}
Brown, R. \& Chua, L.~O. [1998] \enquote{Clarifying chaos {II}: {Bernoulli}
  chaos, zero {Lyapunov} exponents and strange attractors,} \emph{Int. J.
  Bifurcation Chaos} \textbf{8},  1--32, \doi{10.1142/S0218127498000024}.

\bibitem[{Brown \& Chua(1999)}]{cb99}
Brown, R. \& Chua, L.~O. [1999] \enquote{Clarifying chaos {III}: Chaotic and
  stochastic processes, chaotic resonance, and number theory,} \emph{Int. J.
  Bifurcation Chaos} \textbf{9},  785--803, \doi{10.1142/S0218127499000560}.

\bibitem[{Burton(2007)}]{b07}
Burton, D. [2007] \emph{Elementary Number Theory}, 6th ed. (McGraw-Hill, New
  York).

\bibitem[{Buzzi(2001)}]{b01}
Buzzi, J. [2001] \enquote{Piecewise isometries have zero topological entropy,}
  \emph{Ergod. Th. \& Dynam. Sys.} \textbf{21},  1371--1377,
  \doi{10.1017/S0143385701001651}.

\bibitem[{Chella \& Ottino(1984)}]{co84}
Chella, R. \& Ottino, J.~M. [1984] \enquote{Conversion and selectivity
  modifications due to mixing in unpremixed reactors,} \emph{Chem. Eng. Sci.}
  \textbf{39},  551--567, \doi{10.1016/0009-2509(84)80052-4}.

\bibitem[{Christov(2011)}]{c11}
Christov, I.~C. [2011] \enquote{From streamline jumping to strange eigenmodes
  and three-dimensional chaos: A tour of the mathematical aspects of granular
  mixing in rotating tumblers,}  PhD thesis, Northwestern University, Evanston,
  Illinois.

\bibitem[{Christov \emph{et~al.}(2011)Christov, Lueptow \& Ottino}]{clo11}
Christov, I.~C., Lueptow, R.~M. \& Ottino, J.~M. [2011] \enquote{Stretching and
  folding versus cutting and shuffling: An illustrated perspective on mixing
  and deformations of continua,} \emph{Am. J. Phys.} \textbf{74},  359--367,
  \doi{10.1119/1.3533213}.

\bibitem[{Christov \emph{et~al.}(2010{\natexlab{a}})Christov, Ottino \&
  Lueptow}]{col10b}
Christov, I.~C., Ottino, J.~M. \& Lueptow, R.~M. [2010{\natexlab{a}}]
  \enquote{Chaotic mixing via streamline jumping in quasi-two-dimensional
  tumbled granular flows,} \emph{Chaos} \textbf{20},  023102,
  \doi{10.1063/1.3368695}.

\bibitem[{Christov \emph{et~al.}(2010{\natexlab{b}})Christov, Ottino \&
  Lueptow}]{col10}
Christov, I.~C., Ottino, J.~M. \& Lueptow, R.~M. [2010{\natexlab{b}}]
  \enquote{Streamline jumping: A mixing mechanism,} \emph{Phys. Rev. E}
  \textbf{81},  046307, \doi{10.1103/PhysRevE.81.046307}.

\bibitem[{Chua \& Lin(1988)}]{cl88}
Chua, L.~O. \& Lin, T. [1988] \enquote{Chaos in digital filters,} \emph{IEEE
  Trans. Circuits Syst.} \textbf{35},  648--658, \doi{10.1109/31.1802}.

\bibitem[{Clifford \emph{et~al.}(1999)Clifford, Cox \& Roberts}]{ccr99}
Clifford, M.~J., Cox, S.~M. \& Roberts, E. P.~L. [1999] \enquote{Reaction and
  diffusion in a lamellar structure: the effect of the lamellar arrangement
  upon yield,} \emph{Physica A} \textbf{262},  294--306,
  \doi{10.1016/S0378-4371(98)00423-3}.

\bibitem[{Clifford \emph{et~al.}(2000)Clifford, Cox \& Roberts}]{ccr00}
Clifford, M.~J., Cox, S.~M. \& Roberts, E. P.~L. [2000] \enquote{The influence
  of a lamellar structure upon the yield of a chemical reaction,} \emph{Trans.
  {IChemE} A} \textbf{78},  371--377, \doi{10.1205/026387600527491}.

\bibitem[{Deane(2006)}]{d06}
Deane, J. H.~B. [2006] \enquote{Piecewise isometries: Applications in
  engineering,} \emph{Meccanica} \textbf{41},  241--252,
  \doi{10.1007/s11012-005-5895-3}.

\bibitem[{Devaney(2003)}]{d03}
Devaney, R.~L. [2003] \emph{An Introduction to Chaotic Dynamical Systems}, 2nd
  ed. (Westview Press, Cambridge, MA).

\bibitem[{di~Bernardo \emph{et~al.}(2008)di~Bernardo, Budd, Champneys,
  Kowalczyk, Nordmark, Tost \& Piiroinen}]{bbcknop08}
di~Bernardo, M., Budd, C.~J., Champneys, A.~R., Kowalczyk, P., Nordmark, A.~B.,
  Tost, G.~O. \& Piiroinen, P.~T. [2008] \enquote{Bifurcations in nonsmooth
  dynamical systems,} \emph{SIAM Rev.} \textbf{50},  629--701,
  \doi{10.1137/050625060}.

\bibitem[{Franjione \& Ottino(1987)}]{fo87}
Franjione, J.~G. \& Ottino, J.~M. [1987] \enquote{Feasibility of numerical
  tracking of material lines and surfaces in chaotic flows,} \emph{Phys.
  Fluids} \textbf{30},  3641--3643, \doi{10.1063/1.866449}.

\bibitem[{Fu \& Duan(2008)}]{fd08}
Fu, X. \& Duan, J. [2008] \enquote{On global attractors for a class of
  nonhyperbolic piecewise affine maps,} \emph{Physica D} \textbf{237},
  3369--3376, \doi{10.1016/j.physd.2008.07.012}.

\bibitem[{Goetz(1998)}]{g98}
Goetz, A. [1998] \enquote{Dynamics of a piecewise rotation,} \emph{Discret.
  Contin. Dyn. Syst. A} \textbf{4},  593--608, \doi{10.3934/dcds.1998.4.593}.

\bibitem[{Goetz(2000)}]{g00}
Goetz, A. [2000] \enquote{Dynamics of piecewise isometries,} \emph{Illinois J.
  Math.} \textbf{44},  465--478.

\bibitem[{Goetz(2002)}]{g02}
Goetz, A. [2002] \enquote{Piecewise isometries --- an emerging area of
  dynamical systems,}  \emph{Fractals in Graz 2001}, eds. Grabner, P. \& Woess,
  W. (Birkh\"auser, Basel), pp. 135--144.

\bibitem[{Golomb(1961)}]{g61}
Golomb, S.~W. [1961] \enquote{Permutations by cutting and shuffling,}
  \emph{{SIAM} Rev.} \textbf{3},  293--297, \doi{10.1137/1003059}.

\bibitem[{Gouillart \emph{et~al.}(2009)Gouillart, Dauchot, Thiffeault \&
  Roux}]{gdtr09}
Gouillart, E., Dauchot, O., Thiffeault, J.-L. \& Roux, S. [2009]
  \enquote{Open-flow mixing: Experimental evidence for strange eigenmodes,}
  \emph{Phys. Fluids} \textbf{21},  023603, \doi{10.1063/1.3080680}.

\bibitem[{Haller(1981)}]{h81}
Haller, H. [1981] \enquote{Rectangle exchange transformations,} \emph{Monatsh.
  Math.} \textbf{91},  215--232, \doi{10.1007/BF01301789}.

\bibitem[{Hmili(2010)}]{hmli10}
Hmili, H. [2010] \enquote{Non topologically weakly mixing interval exchanges,}
  \emph{Discret. Contin. Dyn. Syst. A} \textbf{27},  1079--1091,
  \doi{10.3934/dcds.2010.27.1079}.

\bibitem[{Hobbs \& Muzzio(1997)}]{hm97}
Hobbs, D.~M. \& Muzzio, F.~J. [1997] \enquote{The {Kenics} static mixer: a
  three-dimensional chaotic flow,} \emph{Chem. Eng. J.} \textbf{67},  153--166,
  \doi{10.1016/S1385-8947(97)00013-2}, see also
  \url{http://www.chemineer.com/kenics_km_static_mixers.php}.

\bibitem[{Jeffrey(2011)}]{j11}
Jeffrey, M.~R. [2011] \enquote{Nondeterminism in the limit of nonsmooth
  dynamics,} \emph{Phys. Rev. Lett.} \textbf{106},  254103,
  \doi{10.1103/PhysRevLett.106.254103}.

\bibitem[{Juarez \emph{et~al.}(2012)Juarez, Christov, Ottino \&
  Lueptow}]{jcol12}
Juarez, G., Christov, I.~C., Ottino, J.~M. \& Lueptow, R.~M. [2012]
  \enquote{Mixing by cutting and shuffling {3D} granular flow in spherical
  tumblers,} \emph{Chem. Eng. Sci.} \textbf{73},  195--207,
  \doi{10.1016/j.ces.2012.01.044}.

\bibitem[{Juarez \emph{et~al.}(2010)Juarez, Lueptow, Ottino, Sturman \&
  Wiggins}]{jlosw10}
Juarez, G., Lueptow, R.~M., Ottino, J.~M., Sturman, R. \& Wiggins, S. [2010]
  \enquote{Mixing by cutting and shuffling,} \emph{EPL} \textbf{91},  20003,
  \doi{10.1209/0295-5075/91/20003}.

\bibitem[{Kahng(2002)}]{k02}
Kahng, B. [2002] \enquote{Dynamics of symplectic piecewise affine elliptic
  rotation maps on tori,} \emph{Ergod. Th. \& Dynam. Sys.} \textbf{22},
  483--505, \doi{10.1017/S0143385702000238}.

\bibitem[{Kahng(2009{\natexlab{a}})}]{k09}
Kahng, B. [2009{\natexlab{a}}] \enquote{On {Devaney}'s definition of chaos for
  discontinuous dynamical systems,}  \emph{Recent Advances in Applied
  mathematics and Computational Information Sciences}, eds. Jegdic, K.,
  Simeonov, P. \& Zafiris, V. (WSEAS Press, Athens), pp. 89--94.

\bibitem[{Kahng(2009{\natexlab{b}})}]{k09b}
Kahng, B. [2009{\natexlab{b}}] \enquote{Singularities of two-dimensional
  invertible piecewise isometric dynamics,} \emph{Chaos} \textbf{19},  023115,
  \doi{10.1063/1.3119464}.

\bibitem[{Katok(1980)}]{k80}
Katok, A. [1980] \enquote{Interval exchange transformations and some special
  flows are not mixing,} \emph{Israel J. Math.} \textbf{35},  301--310,
  \doi{10.1007/BF02760655}.

\bibitem[{Katok \& Hasselblatt(1995)}]{kh95}
Katok, A. \& Hasselblatt, B. [1995] \emph{Introduction to the Modern Theory of
  Dynamical Systems}, Encyclopedia of Mathematics and Its Applications, Vol.~54
  (Cambridge University Press, Cambridge).

\bibitem[{Keane(1975)}]{k75}
Keane, M. [1975] \enquote{Interval exchange transformations,} \emph{Math. Z.}
  \textbf{141},  25--31, \doi{10.1007/BF01236981}.

\bibitem[{Keane(1977)}]{k77}
Keane, M. [1977] \enquote{Non-ergodic interval exchange transformations,}
  \emph{Israel J. Math.} \textbf{26},  188--196, \doi{10.1007/BF03007668}.

\bibitem[{Klazar(2003)}]{klaz}
Klazar, M. [2003] \enquote{Irreducible and connected permutations,} ITI Series
  Preprint 122, Charles University, Prague,
  \urlprefix\url{http://kam.mff.cuni.cz/~klazar/irre.pdf}.

\bibitem[{Liang \& West(2008)}]{lw08}
Liang, T.-C. \& West, M. [2008] \enquote{Numerical evidence for cutoffs in
  chaotic microfluidic mixing,}  \emph{Proceedings of ASME 2008 Dynamic Systems
  and Control Conference} (ASME, New York, NY), pp. 1405--1412,
  \doi{10.1115/DSCC2008-2293}.

\bibitem[{Lowenstein \& Vivaldi(2010)}]{lv10}
Lowenstein, J.~H. \& Vivaldi, F. [2010] \enquote{Approach to a rational
  rotation number in a piecewise isometric system,} \emph{Nonlinearity}
  \textbf{23},  2677--2721, \doi{10.1088/0951-7715/23/10/017}.

\bibitem[{Martelli \emph{et~al.}(1998)Martelli, Dang \& Seph}]{mds98}
Martelli, M., Dang, M. \& Seph, T. [1998] \enquote{Defining chaos,} \emph{Math.
  Mag.} \textbf{71},  112--122,
  \urlprefix\url{http://www.jstor.org/stable/2691012}.

\bibitem[{Masur(1982)}]{m82}
Masur, H. [1982] \enquote{Interval exchange transformations and measured
  foliations,} \emph{Ann. Math.} \textbf{115},  168--200,
  \urlprefix\url{http://www.jstor.org/stable/1971341}.

\bibitem[{Mathew \emph{et~al.}(2005)Mathew, Mezi{\'c} \& Petzold}]{mmp05}
Mathew, G., Mezi{\'c}, I. \& Petzold, L. [2005] \enquote{A multiscale measure
  for mixing,} \emph{Physica D} \textbf{211},  23--46,
  \doi{10.1016/j.physd.2005.07.017}.

\bibitem[{Meier \emph{et~al.}(2007)Meier, Lueptow \& Ottino}]{mlo07}
Meier, S.~W., Lueptow, R.~M. \& Ottino, J.~M. [2007] \enquote{A dynamical
  systems approach to mixing and segregation of granular materials in
  tumblers,} \emph{Adv. Phys.} \textbf{56},  757--827,
  \doi{10.1080/00018730701611677}.

\bibitem[{Mendes \& Nicol(2004)}]{mn04}
Mendes, M. \& Nicol, M. [2004] \enquote{Periodicity and recurrence in piecewise
  rotations of {Euclidean} spaces,} \emph{Int. J. Bifurcation Chaos}
  \textbf{14},  2353--2361, \doi{10.1142/S0218127404010813}.

\bibitem[{Metcalfe \emph{et~al.}(2010)Metcalfe, Lester, Ord, Kulkarni, Trefry,
  Hobbs, Regenaur-Lieb \& Morris}]{mlokthrlm10}
Metcalfe, G., Lester, D., Ord, A., Kulkarni, P., Trefry, M., Hobbs, B.~E.,
  Regenaur-Lieb, K. \& Morris, J. [2010] \enquote{A partially open porous media
  flow with chaotic advection: towards a model of coupled fields,} \emph{Phil.
  Trans. R. Soc. A} \textbf{386},  217--230, \doi{10.1098/rsta.2009.0198}.

\bibitem[{Muzzio \emph{et~al.}(1992)Muzzio, Meneveau, Swanson \&
  Ottino}]{mmso92}
Muzzio, F.~J., Meneveau, C., Swanson, P.~D. \& Ottino, J.~M. [1992]
  \enquote{Scaling and multifractal properties of mixing in chaotic flows,}
  \emph{Phys. Fluids A} \textbf{4},  1439--1456, \doi{10.1063/1.858419}.

\bibitem[{Muzzio \& Ottino(1989)}]{mo89}
Muzzio, F.~J. \& Ottino, J.~M. [1989] \enquote{Evolution of a lamellar system
  with diffusion and reaction: A scaling approach,} \emph{Phys. Rev. Lett.}
  \textbf{63},  47--50, \doi{10.1103/PhysRevLett.63.47}.

\bibitem[{Neerincx \emph{et~al.}(2011)Neerincx, Denteneer, Peelen \&
  Meijer}]{ndpm11}
Neerincx, P.~E., Denteneer, R. P.~J., Peelen, S. \& Meijer, H. E.~H. [2011]
  \enquote{Compact mixing using multiple splitting, stretching, and recombining
  flows,} \emph{Macromol. Mater. Eng.} \textbf{296},  349--361,
  \doi{10.1002/mame.201000338}.

\bibitem[{Novak(2009)}]{n09}
Novak, C.~F. [2009] \enquote{Discontinuity-growth of interval-exchange maps,}
  \emph{J. Mod. Dynam.} \textbf{3},  379--405, \doi{10.3934/jmd.2009.3.379}.

\bibitem[{{OEIS Foundation Inc.}(2012)}]{OEIS}
{OEIS Foundation Inc.} [2012] \enquote{The on-line encyclopedia of integer
  sequences,} \urlprefix\url{http://oeis.org/A003319}.

\bibitem[{Ogorza{\l}ek(1992)}]{o92}
Ogorza{\l}ek, M.~J. [1992] \enquote{Complex behavior in digital filters,}
  \emph{Int. J. Bifurcation Chaos} \textbf{2},  11--29,
  \doi{10.1142/S0218127492000033}.

\bibitem[{Ottino(1989)}]{o89}
Ottino, J.~M. [1989] \emph{The Kinematics of Mixing: Stretching, Chaos, and
  Transport} (Cambridge University Press, Cambridge), \doi{10.2277/0521368782}.

\bibitem[{Ottino \emph{et~al.}(1992)Ottino, Muzzio, Tjahjadi, Franjione, Jana
  \& Kusch}]{omtfjk92}
Ottino, J.~M., Muzzio, F.~J., Tjahjadi, M., Franjione, J.~G., Jana, S.~C. \&
  Kusch, H.~A. [1992] \enquote{Chaos, symmetry, and self-similarity: Exploiting
  order and disorder in mixing processes,} \emph{Science} \textbf{257},
  754--760, \doi{10.1126/science.257.5071.754}.

\bibitem[{Pikovsky \& Rosenblum(2007)}]{pr07}
Pikovsky, A. \& Rosenblum, M. [2007] \enquote{Synchronization,}
  \emph{Scholarpedia} \textbf{2},  1459, \doi{10.4249/scholarpedia.1459},
  \url{http://www.scholarpedia.org/article/Arnold_tongue}.

\bibitem[{Scott \emph{et~al.}(2001)Scott, Holmes \& Milburn}]{shm01}
Scott, A.~J., Holmes, C.~A. \& Milburn, G.~J. [2001] \enquote{{Hamiltonian}
  mappings and circle packing phase spaces,} \emph{Physica D} \textbf{155},
  34--50, \doi{10.1016/S0167-2789(01)00263-9}.

\bibitem[{Sturman(2012)}]{sturman12}
Sturman, R. [2012] \enquote{The role of discontinuities in mixing,} \emph{Adv.
  Appl. Mech.} \textbf{45},  51--90, \doi{10.1016/B978-0-12-380876-9.00002-1}.

\bibitem[{Sturman \emph{et~al.}(2008)Sturman, Meier, Ottino \&
  Wiggins}]{smow08}
Sturman, R., Meier, S.~W., Ottino, J.~M. \& Wiggins, S. [2008] \enquote{Linked
  twist map formalism in two and three dimensions applied to mixing in tumbled
  granular flows,} \emph{J. Fluid Mech.} \textbf{602},  129--174,
  \doi{10.1017/S002211200800075X}.

\bibitem[{Sturman \emph{et~al.}(2006)Sturman, Ottino \& Wiggins}]{sow06}
Sturman, R., Ottino, J.~M. \& Wiggins, S. [2006] \emph{The Mathematical
  Foundations of Mixing}, Cambridge Monographs on Applied and Computational
  Mathematics, Vol.~22 (Cambridge University Press, Cambridge),
  \doi{10.2277/0521868130}.

\bibitem[{Thiffeault(2012)}]{jlt12}
Thiffeault, J.-L. [2012] \enquote{Using multiscale norms to quantify mixing and
  transport,} \emph{Nonlinearity} \textbf{25},  R1--R44,
  \doi{10.1088/0951-7715/25/2/R1}.

\bibitem[{Toussaint \& Carri\`{e}re(1999)}]{tc99}
Toussaint, V. \& Carri\`{e}re, P. [1999] \enquote{Diffusive cut-off scale of
  fractal surfaces in chaotic mixing,} \emph{Int. J. Bifurcation Chaos}
  \textbf{9},  443--454, \doi{10.1142/S0218127499000298}.

\bibitem[{Trefethen \& Trefethen(2000)}]{tt00}
Trefethen, L.~N. \& Trefethen, L.~M. [2000] \enquote{How many shuffles to
  randomize a deck of cards?} \emph{Proc. R. Soc. Lond. A} \textbf{456},
  2561--2568, \doi{10.1098/rspa.2000.0625}.

\bibitem[{Veech(1982)}]{v82}
Veech, W.~A. [1982] \enquote{Gauss measures for transformations on the space of
  interval exchange maps,} \emph{Ann. Math.} \textbf{115},  201--242,
  \urlprefix\url{http://www.jstor.org/stable/1971391}.

\bibitem[{Viana(2006)}]{v06}
Viana, M. [2006] \enquote{Ergodic theory of interval exchange maps,} \emph{Rev.
  Mat. Complut.} \textbf{19},  7--100.

\bibitem[{Walters(1982)}]{w82}
Walters, P. [1982] \emph{An Introduction to Ergodic Theory}, Graduate Texts in
  Mathematics, Vol.~79 (Springer-Verlag, New York).

\end{thebibliography}

\end{document}